\documentclass[twocolumn]{aastex631}
\usepackage{amsmath}
\usepackage{enumitem}   
\usepackage{natbib}
\defcitealias{drout_2023}{DGL23}
\defcitealias{ludwig_2025}{L25}

\begin{document}

\title{Intermediate-Mass Stripped Stars in the Magellanic Clouds: Forward Modeling the Observed Population Discovered Via UV Excess}

\correspondingauthor{lblomber@caltech.edu}

\author{Lisa Blomberg}
\affiliation{Department of Astronomy, California Institute of Technology, 1200 East California Boulevard, Pasadena, CA 91125, USA}
\author{Kareem El-Badry}
\affiliation{Department of Astronomy, California Institute of Technology, 1200 East California Boulevard, Pasadena, CA 91125, USA}
\author{Bethany Ludwig}
\affiliation{Institute of Astronomy, KU Leuven, Celestijnenlaan 200D, 3001, Leuven, Belgium}
\author{Maria R. Drout}
\affiliation{The David A. Dunlap Department of Astronomy \& Astrophysics, University of Toronto, 50 St. George Street, Toronto, Ontario, M5S 3H4, Canada}
\author{Ylva G\"{o}tberg}
\affiliation{Institute of Science and Technology Austria (ISTA), Am Campus 1, 3400 Klosterneuburg, Austria}





\begin{abstract}
Stripped stars are hot, helium-rich stars formed when binary interactions remove a star's hydrogen envelope. While low-mass ($\lesssim 1\,M_\odot$) and high-mass ($\gtrsim 8\,M_\odot$) stripped stars are well studied as hot subdwarfs and Wolf-Rayet stars, their intermediate-mass counterparts ($1-8 \,M_\odot$) have only recently been discovered. The Stripped-Star Ultraviolet Magellanic Cloud Survey (SUMS) identified UV-excess sources (i.e., sources lying blueward of the main sequence) in the Magellanic Clouds using \textit{Swift}-UVOT photometry and selected 820 photometric stripped-star candidates. However, the completeness and purity of this sample remain poorly understood. We forward model the population of stripped stars in the Magellanic Clouds using a binary population synthesis model combined with spatially resolved star formation histories and simulated UV photometry. To assess survey sensitivity, we inject simulated sources into real \textit{Swift}-UVOT images and reproduce the SUMS selection process, including crowding, extinction, and photometric quality cuts. For the SMC and LMC, we respectively recover 31\% and 15\% of sources with intrinsic UV excess, and $11\%$ and $7\%$ of all stripped stars. The rest are missed due to dilution by luminous companions, crowding, high extinction, and limited survey coverage. The observed population is biased toward systems with low-mass companions formed by common envelope evolution and systems with compact object companions. We predict contamination of the observed stripped-star candidates by main-sequence stars with spurious UV excess due to crowding and provide guidelines for selecting higher-purity subsamples. Our population synthesis model matches the number and properties of observed stripped star candidates with masses of $(2-5)\,M_\odot$ well. It underpredicts the number of lower- and higher-mass candidates, perhaps due to contamination.
\end{abstract}

\keywords{}


\section{Introduction} \label{sec:intro}

Stars in binary systems can be stripped of their hydrogen-rich envelopes -- through mass transfer or common envelope (CE) ejection -- leaving behind stripped helium stars \citep{Kippenhahn_1967, Paczyski_1967, Paczynski_1976}. Low-mass ($\lesssim 1\,M_\odot$) and high-mass ($\gtrsim 8 \,M_\odot$) stripped stars are called hot subdwarf (sdO/B) stars and Wolf-Rayet (WR) stars, respectively, and have been studied extensively over the last several decades 
\citep[e.g.,][]{crowther_2007, heber_2016}. Intermediate-mass stripped stars, with masses of $(1-8) \,M_\odot$, are predicted to form through similar binary formation channels as sdO/B and WR stars, but have until recently eluded observational searches (\citealt{drout_2023}, hereafter \citetalias{drout_2023}).

Intermediate-mass stripped stars are formed via binary envelope stripping of stars with initial masses of $(5-25)\,M_{\odot}$ \citep{Gotberg_2018}. They have typical lifetimes of order $(1-50)$ Myr \citep{hovis_2025} and are predicted to be abundant. Binary population synthesis models estimate $\sim 8 \times 10^3$ and $\sim 6 \times 10^4$ binaries in the Milky Way (MW) containing stripped stars more massive than $2.6\,M_\odot$ and $1\,M_\odot$, respectively \citep{hovis_2025, Yungelson_2024}. Similarly, they are predicted to be common in the Magellanic Clouds (MCs), with population synthesis models estimating $\sim6,000$ and $\sim1,600$ binaries containing stripped stars more massive than $1\,M_\odot$ in the Large Magellanic Clouds (LMC) and Small Magellanic Cloud (SMC), respectively. However, only a small number of intermediate-mass stripped stars have been discovered so far, due to the difficulty of distinguishing them in photometric surveys from more numerous main-sequence (MS) stars and other blue objects. Because stripped stars are hot ($T_{\rm eff} \gtrsim 50{,}000$ K; \citealt{Gotberg_2018}), they are bright in the ultraviolet (UV). Most stripped stars are predicted to be in binaries with MS companions \citep{pols_1994}. Depending on the mass ratio, a subset of these systems are expected to be brighter in the UV than single MS stars with similar visual magnitudes \citep{Gotberg_2018}.

The first sample of hot intermediate-mass stripped stars was discovered in the MCs by searching for systems with excess UV light using new UV photometry performed on images from the \textit{Swift} Ultraviolet/Optical Telescope \citep[UVOT; ][]{Gehrels_2004, swift_UVOT, Poole_2008}. The Stripped-Star Ultraviolet Magellanic Cloud Survey (SUMS; \citealt{ludwig_2025}; hereafter L25) identified $\sim 800$ photometric candidates that lie bluewards of the Zero-Age Main Sequence (ZAMS) in UV-optical color-magnitude diagrams (CMDs). 16 candidates from a preliminary version of the survey (4 in the SMC and 12 in the LMC) have been spectroscopically confirmed \citepalias{drout_2023}.

The population demographics (e.g., occurrence rates, mass distribution, binary fraction, and orbital period distribution) of stripped stars are still poorly constrained, in part because the purity and completeness of the SUMS candidates have not yet been characterized. In particular, \citetalias{drout_2023} highlight that for an intermediate-mass stripped star, detectable UV excess is expected only if $M_{\rm MS} / M_{\rm stripped} < 1.4 - 2.5$. This technique therefore preferentially selects systems that have undergone non-conservative mass transfer or common envelope evolution (CEE). However, full quantification of the selection function and sensitivity of the survey has not yet been performed.

Observational constraints on the population can provide powerful tests of binary evolution models \citep[e.g.][]{shao_2021, Yungelson_2024, Wang2024}. Binaries containing stripped stars are progenitors of several classes of astrophysically interesting objects, such as double neutron stars \citep{Tauris_2017} and stripped-envelope supernovae \citep[e.g., ][]{smith_2011, Drout_2011, eldridge_2013, lyman_2016}. Additionally, the high temperatures of stripped stars make them potentially important sources of ionizing photons during cosmic reionization \citep[e.g.,][]{stanway_2016, gotberg_2020}. Understanding the population demographics of stripped stars and the binary interaction processes that produce them can thus inform models for a wide range of astrophysical phenomena.

In this paper, we forward model the UV photometry and stripped-star candidate selection of the SUMS catalog. We modify the publicly available photometric pipeline used by \citetalias{ludwig_2025} to perform source injection, of mock stripped stars derived from a synthetic population into the UVOT data, and assess the recoverability of stripped stars by the survey. By assessing the synthetic population of stripped stars missing from a UV-excess sample, we quantify the incompleteness of the SUMS candidate population and infer the underlying population of stripped stars in the MCs. We create a synthetic population of stripped-star binaries by Monte Carlo sampling from a binary population synthesis model based on binary evolutionary tracks anchored to spatially-resolved star formation histories of the SMC and LMC. We model a realistic 3D dust distribution of the Clouds to compute reddening for each synthetic source. To model UV photometry, we inject synthetic sources into the \textit{Swift}-UVOT images used in the observational search and perform mock photometry following the same methods used to construct the SUMS catalog. Our model accounts for reddening due to dust, dilution of the stripped stars' light by their companions, blending of the stripped stars' light with nearby sources in crowded regions, and the finite spatial coverage of SUMS.

The rest of the paper is organized as follows. In Section \ref{sec: observed systems}, we summarize the observed population of stripped stars in the SUMS catalog. In Section \ref{sec: forward modeling}, we describe forward modeling the stripped-star population in the MCs, including creating a synthetic population of stripped stars, modeling dust extinction, injecting them into \textit{Swift}-UVOT images, performing forced photometry, and selecting systems with robust UV excess. In Section \ref{sec: mock observed population}, we compare the mock-observed population to the observed stripped-star candidates and predict the complete population of stripped stars in the MCs. We also test the purity of the observed sample by forward modeling a population of MS stars to test their contamination rate in a UV-excess survey. Finally, we conclude with a discussion of our findings and their implications in Section \ref{sec: conclusion}.

\section{observed population}\label{sec: observed systems}

SUMS \citepalias{ludwig_2025} carried out a UV-excess survey in the MCs and produced a catalog of 820 photometric stripped-star candidates. Of which, a subset from a preliminary version of the survey have been spectroscopically confirmed \citepalias{drout_2023}. Below, we summarize the survey data, selection, and the properties of the spectroscopically-confirmed sources.

\subsection{Data and selection function}

\subsubsection{Photometry}
The SUMS catalog used a combination of optical and UV photometry to select UV-excess sources.

\textit{Optical photometry: }optical photometry was taken from the Magellanic Cloud Photometric Survey (MCPS), which was conducted using the Great Circle Camera \citep{zaritsky_1996} on the $1\,{\rm m}$ Swope telescope at Las Campanas Observatory.  The survey covers $4.5^\circ\times4^\circ$ in the SMC and $8.5^\circ\times7.5^\circ$ in the LMC, and includes over 5 million sources in the SMC and 24 million sources in the LMC, with optical photometry in the $U$, $B$, $V$, and $I$ bands. 

\textit{UV photometry: } between 2010 and 2013, the \textit{Swift}-Ultraviolet/Optical Telescope \citep[UVOT; ][]{Gehrels_2004, swift_UVOT, Poole_2008} carried out the \textit{Swift} Ultraviolet Survey of the Magellanic Clouds \citep[SUMaC; ][]{Siegel_2015, hagen_2016}. These observations consist of 150 fields in the LMC and 50 fields in the SMC. \citetalias{ludwig_2025} computed UV photometry for $UVW2$, $UVM2$, and $UVW1$ bands from the \textit{Swift}-UVOT images using \textit{The Tractor} \citep{Lang_2016}, and measured UV photometry for 734,869 sources (see Section 3 of \citetalias{ludwig_2025} for details). 

\subsubsection{Extinction}
SUMS de-reddened their observed photometry using a single average extinction value for each Cloud, rather than modeling reddening for individual sources. They adopt $A_V = 0.22$ mag for the SMC and $A_V = 0.38$ mag for the LMC, such that the theoretical ZAMS lies along the blue edges of the MS in UV-optical CMDs (See S3.5 of \citetalias{drout_2023} and Section 5.2 of \citetalias{ludwig_2025} for further discussion). \citetalias{ludwig_2025} noted that this choice will exclude true stripped-star binaries in regions of high extinction from their sample, but considered this preferable to significant contamination by MS stars with over-corrected extinction.



\subsubsection{Selecting stripped-star candidates}
SUMS identifies UV-excess sources by comparing their data with a theoretical ZAMS, constructed as described in S1.3.2 of \citetalias{drout_2023}. 
For sources that show robust UV excess relative to the theoretical ZAMS, they apply a cascade of cuts, including photometric and SED quality checks, and cuts related to the number of nearby sources in the \textit{Swift}-UVOT images, to minimize sources with spurious UV photometry (Section 5 of \citetalias{ludwig_2025}). We will further describe this selection in Section \ref{sec: candidate selection}.

\subsection{The SUMS catalog}
\begin{figure*}
    \centering
    \includegraphics[width=\linewidth]{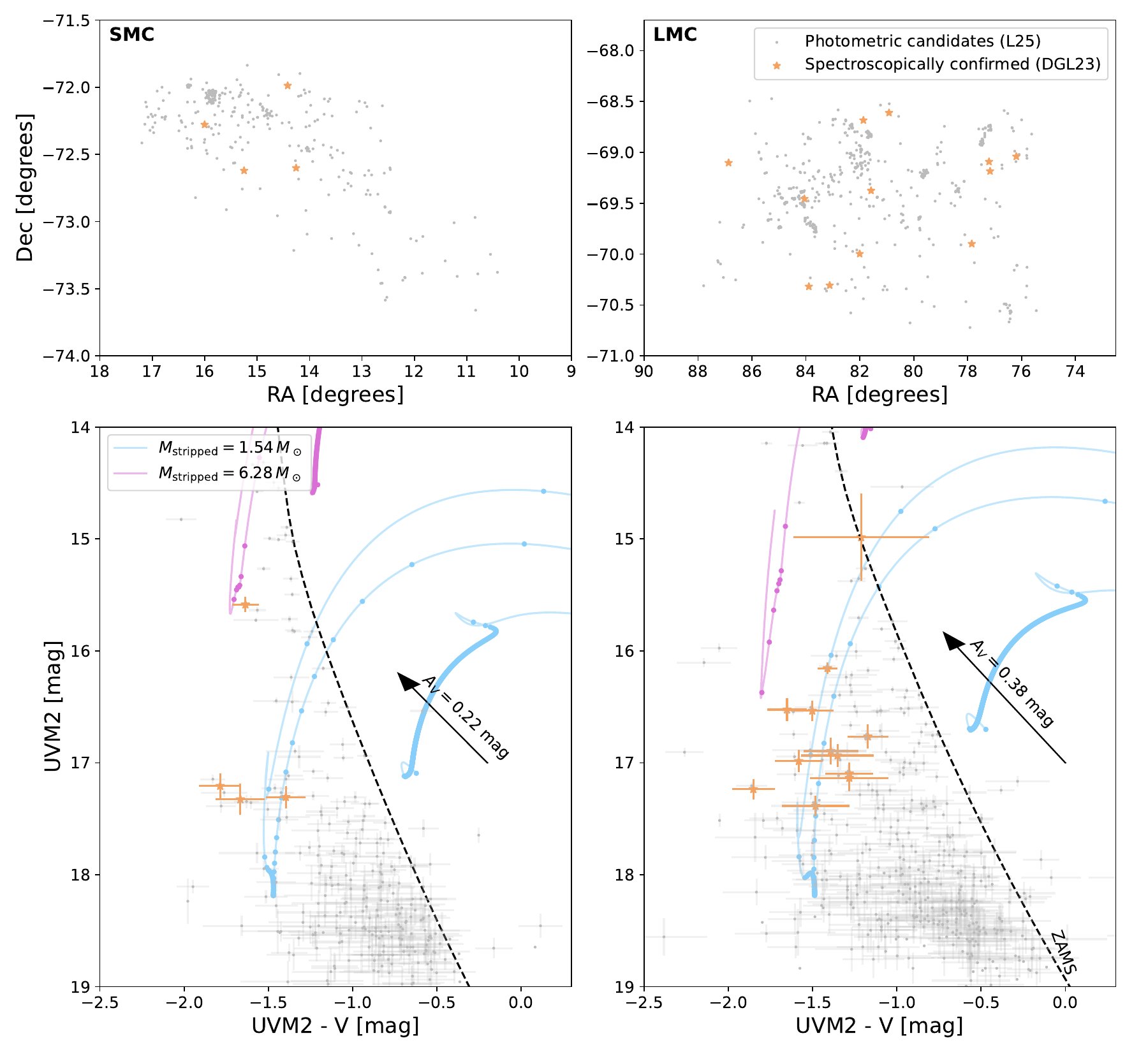}
    \caption{Observed stripped-star candidates in the SMC (left) and LMC (right). Top: spatial distribution of the spectroscopically-confirmed sources \citepalias[orange;][]{drout_2023} and photometric candidates \citepalias[gray;][]{ludwig_2025}. Bottom: one of the nine UV-optical CMDs used to assess UV excess. UV and optical photometry is de-reddened assuming a uniform extinction of $A_V = 0.22$ mag for the SMC and $A_V = 0.38$ mag for the LMC. Dashed line indicates the theoretical ZAMS. Purple and blue lines show evolutionary tracks for stripped stars with masses of $M_{\rm stripped} = 6.28\,M_\odot$ ($M_{\rm stripped, init} = 16.44\,M_\odot$) and $M_{\rm stripped} = 1.54\,M_\odot$ ($M_{\rm stripped, init} = 6.03\,M_\odot$), respectively. Points along each track are spaced $10^{5}$ years apart. These tracks  show only the photometry from the stripped star and do not account for flux contributions from the companion.}
    \label{fig: drout cmd}
\end{figure*}

The SUMS catalog contains 820 photometric stripped-star candidates, which are divided into four categories (Section 5.7 of \citetalias{ludwig_2025}) based on (i) the strength of their detected UV excess (``Very Blue'' or ``Blue'') and (ii) their SED and photometric quality (“Excellent” or “Good”). 
Figure \ref{fig: drout cmd} shows a UV-optical CMD and spatial distribution of the candidates. The stripped-star candidates identified in \citetalias{ludwig_2025} encompass a broader range of magnitudes than the spectroscopically confirmed stripped stars from \citetalias{drout_2023}.

\subsubsection{Spectroscopically-confirmed systems}
\label{sec:spec_confirmed}
42 stripped-star candidates from a preliminary version of the \citetalias{ludwig_2025} catalog were followed up spectroscopically by \citetalias{drout_2023}. 
17 were rejected because they were either: found to not have UV excess with the improved photometric analysis  (9 objects), were classified as likely foreground objects (4 objects), or showed complex spectral morphology inconsistent with theoretical models of stripped-star spectra (4 objects). 25 objects were identified with luminosities, colors, and spectral morphologies consistent with expectations for stripped-star binaries. 8 sources showed optical spectra consistent with being dominated by a hot helium star (``class 1''), 8 sources showed spectroscopic signatures of both a B-type MS star and a hot helium star (``class 2''), and 9 sources showed spectra consistent with a B-type MS star, despite showing an apparent UV excess (``class 3''). \citetalias{drout_2023} demonstrate that all three spectral types are expected for stripped-star binaries, depending on the mass ratio of the stripped star and its MS companion, though they caution that the nature of the class 3 sources remains more uncertain.

Atmospheric properties of the class 1 stripped stars have been studied through spectral modeling, revealing high effective temperatures, $T_{\rm eff} = 50-100$ kK, high surface gravities, $\log(g / \rm{cm\, s}^{-2}) \sim 5$, hydrogen-depleted surfaces, $X_{\rm H}\lesssim 0.4$, and masses between $(2-8)\,M_{\odot}$, which are consistent with the expected properties of intermediate-mass stripped stars \citep{gotberg_2023}.

As discussed in S3.2 of \citetalias{drout_2023}, the study was exploratory and did not select follow-up candidates in a systematic or reproducible way. Many of the 42 candidates selected for spectroscopic follow-up were chosen ``by-eye'' because they appeared particularly promising and do not pass all cuts used by the SUMS catalog (e.g. objects with weaker/less significant UV excess, objects with kinematics that are inconsistent with the MCs). The detection and contamination rates of this spectroscopically-confirmed population are thus unlikely to be representative of the full \citetalias{ludwig_2025} photometric candidate population, even in the same magnitude range. We therefore focus on the \citetalias{ludwig_2025}  photometric candidates, which have a well-understood selection function.

\subsection{Completeness and purity}\label{sec: selection biases}
The population of stripped-star candidates selected via a UV-excess search is expected to be incomplete due to a variety of selection effects. Below, we summarize the most important of these:

\begin{enumerate}[label=(\roman*)]
    \item \textit{Bright companions} can dilute the UV excess of stripped stars, causing them to not show a detectable UV excess (Section \ref{sec: companion discussion}).
    \item \textit{The finite footprint}
    of the SUMaC images used to produce the SUMS catalog limits the search to the densest regions of the Clouds surveyed by \textit{Swift}-UVOT (Section \ref{sec: footprint discussion}). Saturation and coincidence loss artifacts also restrict the spatial area with usable photometry.
    \item \textit{Crowding} can cause the brightness of a source to be degenerate with a nearby source, which can dilute the UV excess of a stripped star (Section \ref{sec: crowding}).
    \item \textit{Dust extinction} can redden stripped-star binaries. This can hide the UV excess in highly extincted regions, causing them to fall on the MS in UV-optical CMDs (Section \ref{sec: extinction discussion}).
\end{enumerate}
In this paper, we characterize the above biases by forward modeling the photometry and selection function of SUMS.

We also assess the purity of the photometric sample. We note that in the \emph{spectroscopic} sample presented by \citetalias{drout_2023}, 4 sources were rejected by the final sample despite showing robust UV excess and kinematics consistent with Cloud membership, because they had spectral morphologies (e.g. metal absorption lines and/or emission lines) that differ from theoretical spectral models of stripped stars.
However, we note that while some of these objects are likely contaminants, the nature of the objects have not yet been confirmed. Additionally, of the 25 stars presented in the final \citetalias{drout_2023} sample, the nine ``Class 3'' objects have spectra similar to normal B-type stars (albeit with a UV excess). While this is expected for a subset of stripped-star binaries, especially those with luminous companions, the nature of individual Class 3 stars are less secure than the Class 1 and 2 sources. As discussed in Section 2.2.1, the contamination rate for the \citetalias{drout_2023} candidates is unlikely to be representative of the SUMS candidates (e.g., 6 out of 9 Class 3 sources from \citetalias{drout_2023} do not make the SUMS photometric candidates list). \citetalias{ludwig_2025} demonstrate that several types of contaminants are expected, including some MS stars, non-members of the Clouds, WR stars, post-AGB stars, planetary nebulae, and rapidly accreting white dwarfs. They show that the most common contaminants are likely MS stars, which may enter the candidate sample in two ways:
\begin{enumerate}[label=(\roman*)]
    \item Overcorrected extinction: if a MS star lies in a region of particularly low reddening, the photometry could be overcorrected, making the source appear bluer than it is intrinsically. \citetalias{ludwig_2025} demonstrated that this likely only impacts the ``Blue'' subset of their candidate sample.
    \item Crowding: in addition to diluting the true UV excess of a source, crowding can also overestimate the UV flux of a source, which artificially amplifies the UV excess of a MS star, resulting in a false-positive detection.
\end{enumerate}

Understanding the completeness and purity of the sample will allow us to estimate the fraction of stripped stars that remain undetected and predict the properties of the underlying population. The goal of this paper is to forward model the SUMS population, predict the completeness and purity of the UV-excess photometric catalog, and assess broad implications for current population synthesis models of stripped stars. Because most of the photometric candidates have not yet been spectroscopically followed-up, the sample's purity is still uncertain. Spectroscopic follow up of the SUMS candidates is ongoing. This work develops the framework necessary for comparison of observed and simulated samples of stripped stars, and we plan to apply it to spectroscopically-confirmed samples in future work.

\section{Forward modeling a synthetic population}\label{sec: forward modeling}

\subsection{Binary population synthesis}\label{sec: starburst parameters}
To produce a mock population of stripped stars consistent with the star formation histories of the MCs, we first create a synthetic instantaneous starburst population of total initial mass $10^6\,M_\odot$, which is large enough that the initial mass function is well-sampled at the masses relevant to intermediate-mass stripped stars. We then Monte Carlo sample from this population using empirical formation histories to model the population of stripped stars formed in the past 100 Myrs.
We only consider stripped stars with masses between $1-8\,M_\odot$. We choose this mass range for two reasons: (1) to only consider stripped stars which are reliably modeled by the binary evolution models described in Section \ref{sec: mesa grids}, and (2) to keep consistent with the targeted mass range of SUMS. Previous studies have shown that stripped stars start to exhibit the WR phenomenon only at masses above $\sim 11\,M_\odot$ and $\sim 16\,M_\odot$ at SMC and LMC metallicities \citep[e.g., ][]{shenar_2020}. The goal of our study is not to model the stripped-star population up to the WR mass, but rather to model the mass range of $M_{\rm stripped} = 1-8\,M_\odot$ targeted by SUMS. 

There are large uncertainties in binary population synthesis, especially related to mass transfer and CEE, which affect predictions for the number and properties of synthetic stripped stars. In this work, we adopt a single population synthesis model as ground truth for developing our forward-modeling framework. We further discuss the limitation of this assumption in Section \ref{sec: dependence on underlying population}. Below, we describe initial parameters and assumptions of our models, which are also used in \cite{gotberg_2020} and \cite{hovis_2025}. We use a binary population synthesis code that interpolates binary evolution tracks, allowing the effective temperatures and luminosities of stripped stars to be modeled more accurately than in standard BSE-based codes that rely on pure helium star grids.

\subsubsection{Binary evolution models and initial conditions}\label{sec: mesa grids}

We use the grid of binary evolution models for stripped-star binaries from \cite{Gotberg_2018}. Below, we summarize the key assumptions of the models. A full list of input parameters can be found in \cite{Gotberg_2018}. The models were constructed with Modules for Experiment in Stellar Astrophysics \citep[MESA;][]{paxton_2010, paxton_2013, paxton_2015, Paxton_2018, paxton_2019, jermyn_2023}. The grid contains 23 primary masses logarithmically sampled between $2\,M_\odot$ and $18.17\,M_\odot$, where the primary star is defined as the initially more massive star in the binary. The initial mass ratio is fixed to $q = M_{\rm 2,\,init} / M_{\rm 1,\,init} = 0.8$.
The metallicity is set to $Z = 0.006$, which approximately matches the LMC metallicity. 

The models assume Case B mass transfer, which is likely the most common channel for producing stripped stars \citep{Sana_2012}. They do not account for scenarios, such as partial-envelope stripping and nuclear-timescale mass transfer \citep[e.g., ][]{klencki_2022}, which may be important at low metallicity, or case C mass transfer, where stripping begins at a later evolutionary stage \citep[e.g., ][]{ercolino_2024, ercolino_2025}. Partial-envelope stripping and nuclear-timescale mass transfer are predicted to be important only near the top of the stripped-star mass range we consider, and they produce cooler ($T_{\rm eff} \leq 50\,\rm{kK}$) stripped stars than predicted by the \citet{Gotberg_2018} models, which make them less important for UV-excess selected populations. Nonetheless, these modes of stripping should be studied in future work.

The models assume rejuvenation of the companion stars following \cite{Tout_1997}. 
Since the properties of a stripped star depend primarily on the initial primary mass and only weakly on the initial orbital period, mass ratio, and formation channel \citep[e.g.][]{yoon_2017}, we interpolate the grid to predict the properties of all intermediate-mass stripped stars. The temperatures and luminosities of stripped stars do depend somewhat on the thickness of the surface hydrogen layer, which depends on the wind prescription for hot stars. For example, \citet{gilkis_2019} show that using the wind prescription of \citet{Vink2017} produces stripped stars that are $\sim 10\%$ cooler and $\sim 10\%$ more luminous than those produced by the prescription of \citet{Nugis2000}, which is assumed in the \citet{Gotberg_2018} models. While these differences are not entirely negligible, they are likely subdominant to uncertainties associated with crowding and extinction.

Following \cite{Gotberg_2019}, we simulate a starburst population by drawing primary masses from a \cite{kroupa_2001} initial mass function sampled between 0.1 and $100\,M_\odot$. We assign companions to a fraction of these stars according to the mass-dependent binary fraction from \cite{moe_2017}. We determine secondary masses by drawing from a uniform mass ratio distribution between 0.1 and 1 \citep[e.g.,][]{chini_2012, Kiminki_2012, Sana_2012, moe_2017, Shenar_2022}. We sample initial orbital periods from a distribution that follows \cite{opik_1924} for $M_{1,\,\rm{init}}< 15 \,M_\odot$ and \cite{Sana_2012} for $M_{1,\,\rm{init}}>15\,M_\odot$, and set the maximum initial period to $10^{3.7}$ days. We assume circular orbits. We sample from these distributions until a total mass of $10^6\,M_\odot$ is reached.

\subsubsection{Mass transfer and common envelope}

We assume that mass transfer begins when the more massive star overflows its Roche lobe. We treat mass transfer as stable if the initial mass ratio of the binary exceeds a critical value: $q_{\rm crit} = 0.65$ for systems that initiate mass transfer while the donor is a MS star, and $q_{\rm crit} = 0.4$ for systems that initiate mass transfer while the donor is in the Hertzsprung gap (\citealt{demink_2007} and \citealt{Hurley_2002}, respectively). We note that the critical mass ratio for stable mass transfer is uncertain. Some recent studies predict that $q_{\rm crit}$ may be as low as $\sim 0.2$ \citep[e.g., ][]{pavlovskii_2016, marchant_2021, klencki_2021}. A lower $q_{\rm crit}$ would yield a larger number of stripped stars, because it would cause more interacting systems to survive mass transfer. However, stripped stars formed via stable mass transfer typically host more massive MS companions than those formed via CEE. Thus, the number of stripped stars with detectable UV excess will {\it decrease} if mass transfer is more stable. We further consider these effects in Section \ref{sec: dependence on underlying population}.

We assume systems with $q < q_{\rm crit}$ and binaries containing a star with a deep convective envelope will undergo CEE. We model CEE using the $\alpha$ formalism and assume $\lambda_{\rm CE} = 0.5$ and $\alpha_{\rm CE}=1$, which corresponds to the average $\lambda_{\rm CE}$ for stars in the middle of the Hertzsprung gap \citep[e.g.,][]{Izzard_2004}. We only consider systems which initiates CEE on the Hertzsprung gap. If either of the stars' cores initiate mass transfer during CEE, the binary is not likely to survive, and we remove the system from the sample.

In summary, we form stripped-star binaries from the primary star through three main formation channels (see Section \ref{sec: stripped stars with dark companions} for cases where the secondary star is stripped):  mass transfer initiated while the donor is a MS star, mass transfer initiated while the donor is in the Hertzsprung gap, and CEE initiated during the Hertzsprung gap evolution of the donor.

\subsubsection{Population interpolation}

We predict the mass and luminosity of the stripped star from the primary's initial mass by interpolating on MESA evolutionary tracks for stars that undergo stable envelope stripping. We determine masses of the stripped stars by interpolating masses of the progenitor stars from the stellar evolutionary models and taking the masses halfway through central helium burning. We assign physical parameters to the companions, including mass and radius, by interpolating the properties from a grid of 40 MESA single-star models (spaced logarithmically between $2\,M_\odot$ and $100\, M_\odot$) with the same initial parameters as the above stripped star MESA binary models described in Section \ref{sec: mesa grids}.

For high-mass stripped stars, $M_{\rm stripped} \gtrsim 7\,M_\odot$, and low-mass MS companions, $M_{\rm MS} < 2\,M_\odot$, which fall outside our MESA grids, we extrapolate to obtain physical properties of the stars. Given uncertainties in the rest of our binary population synthesis model, the resulting uncertainties are likely subdominant. In our model, high-mass stripped stars, usually host massive MS companions, which prevent them from entering the UV-excess sample regardless of the stripped-star properties. The low-mass companions are usually significantly outshone by the stripped star, making the precise properties of the companion unimportant for detecting UV excess.

\subsubsection{Stripped stars with compact companions}\label{sec: stripped stars with dark companions}
The above binary evolution models only consider the case where the initially more massive star is stripped. However, it is also possible for the initially less massive star to evolve into a stripped star. This occurs when the primary star becomes a white dwarf, a neutron star, or a black hole -- usually after an episode of mass transfer -- and subsequently strips its MS companion of its hydrogen envelope. While such systems are expected to be rare, making up $\sim 7\,\%$ of the total stripped-star binary population \citep{zapartas_2017}, they are important in stripped-star samples selected by UV excess because the compact object companions do not outshine the stripped star. Predicting the properties of stripped stars with compact object companions is challenging because it depends on the uncertain fraction of binaries that survive the first compact object's formation. To account for such systems, we follow the assumptions made in \cite{gotberg_2020b}, which selects a random $7\,\%$ of the stripped stars in the $10^6 \,M_\odot$ starburst population to have compact object companions.

We assume that the second stripped star will have the same mass as the first stripped star formed in the system. We do not attempt to predict masses of the compact objects. We refer to \cite{gotberg_2020b} for further discussion and caveats of the above assumptions.

\subsubsection{Synthetic photometry and final outputs}\label{sec: synthetic photometry and final outputs}

We compute synthetic photometry for stripped stars and their companions using a two-step process. We begin with approximate magnitudes calculated assuming blackbody spectra convolved with the relevant transmission curves from the SVO2 Filter Profile Service \citep{rodrigo_2012, rodrigo_2020, rodrigo_2024}. We then apply corrections to these magnitudes by calculating offsets between blackbody magnitudes and magnitudes calculated for realistic spectral models.

We construct a grid of offsets for stars with a range of effective temperatures and surface gravities, using synthetic spectra from \cite{gotberg_2023}, TLUSTY \citep{Hubeny_1995, lanz_2003, lanz_2007}, and the BOSZ Synthetic Stellar Spectral Library \citep{Bohlin_2017, Meszaros_2024}. For stripped stars, we adopt the \citet{gotberg_2023} models for sources with temperatures $T_{\rm eff} > 30{,}000$ K, and use TLUSTY and BOSZ spectra for cooler stars with $15{,}000\,\mathrm{K} < T_{\rm eff} < 30{,}000\,\mathrm{K}$ and $T_{\rm eff} < 15{,}000\,\mathrm{K}$, respectively. For MS and post-MS stars, we use the TLUSTY and BOSZ models following the same temperature thresholds. When a $T_{\rm eff}$ and $\log(g)$ combination is outside of model grids, we hold $T_{\rm eff}$ constant and round to the nearest available $\log(g)$ model. We interpolate within these grids to compute the offsets relative to the blackbody photometry. This approach allows us to predict synthetic photometry for any combination of stripped star and MS star temperatures and luminosities based on a fine grid of spectral models.

We record the properties of all stripped-star binaries in the $10^6\,M_\odot$ starburst population at 100 time steps logarithmically spaced between 7.9 and 100 Myr.


\subsection{Stripped stars from an extended star formation history} \label{sec: stellar formation rate and syntehtic population}

Using the starburst population described in Section \ref{sec: starburst parameters} and spatially-resolved star formation histories of the Clouds, we create a synthetic population of stripped stars via Monte Carlo sampling.

\subsubsection{Spatially-resolved star formation histories}\label{sec: star formation rate}
\cite{rubele_2018} and \cite{Mazzi_2021} measured the star formation history for 168 and 756 subregions in the SMC and LMC, respectively, using the VISTA survey of the Magellanic Clouds \citep[VMC;][]{VMC}. Each SMC subregion covers 0.143 deg$^2$ and the LMC subregion covers 0.125 deg$^2$. In total, the star formation history map covers $\sim 25$ deg$^2$ in the SMC and $\sim 95$ deg$^2$ in the LMC. \cite{rubele_2018} and \cite{Mazzi_2021} use a CMD reconstruction method and fit two sets of near-infrared CMDs for parameters including the time-dependent star formation rate, age-metallicity relation, distance, and mean reddening. Their models include unresolved binaries. In each spatial region, they constrain the star formation history for $14-16$ age bins spaced approximately logarithmically in time. For both Clouds, there are 3–5 age bins for ages $<100$ Myr. The youngest bin represents roughly the past $<1$ Myr and each successive bin has a width of (0.2–0.5) dex in age.

\subsubsection{Sampling stripped stars}
\begin{figure}
    \centering
    \includegraphics[width=\linewidth]{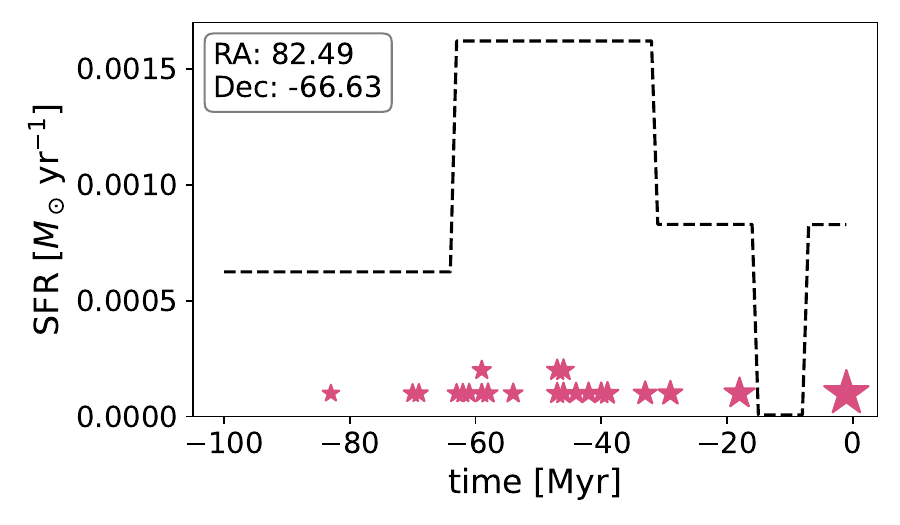}
    \caption{An example star formation history for a LMC subregion, where $t=0$ represents today. The dashed line indicates the star formation rate from \cite{Mazzi_2021}. Star markers show the birth year of stripped stars alive today. The size of the markers corresponds to the mass of the stripped stars at $t = 0$.}
    \label{fig: LMC stripped star formation}
\end{figure}

For each subregion, we compute the total mass $M_i$ formed in each $\Delta t = 1$ Myr interval in the past 100 Myr. Given a star formation rate, $\dot M (t)$, we can predict the mass formed in a given time interval, $i$, as:
\begin{equation}
    M_i = \dot M (t)\Delta t_i.
\end{equation}
We calculate the probability that each stripped star in the $10^6 \,M_\odot$ starburst population evaluated at age $\Delta t_i$ formed during the corresponding time interval $\Delta t_i$:
\begin{equation}\label{eq: star prob}
    P_{\rm stripped}(\Delta t_i) = \frac{M_i}{10^6 M_\odot}.
\end{equation}
For each $\Delta t_i$, we draw a random sample of stripped stars from the starburst at the appropriate age according to Equation \ref{eq: star prob}. If a given $\Delta t_i$ falls between two age snapshots of the starburst population, we sample from the two closest ages while weighting $P_{\rm stripped}$ proportional to their proximity to $\Delta t_i$. We repeat this for all $\Delta t_i$, where $i$ range from 0 to 99, to model the population that formed over the past 100 Myr. We  combine the results to simulate the expected population of stripped stars today in the MCs. Figure \ref{fig: LMC stripped star formation} shows an example stripped star formation history for a single subregion of the LMC. The more massive stripped stars were formed more recently, compared to lower-mass stripped stars, because massive stars have shorter lifetimes.

Our simulated population contains 1,019 (SMC) and 2,486 (LMC) stripped-stars binaries with masses between $1\,M_\odot \leq M_{\rm stripped} \leq 8 \,M_\odot$. Of the 3,505 simulated stripped stars, 3,196 (938 in the SMC and 2,258 in the LMC) host MS companions, 31 (8 in the SMC and 23 in the LMC) host post-MS companions, and 278 (73 in the SMC and 205 in the LMC) host compact object companions. The number of simulated stripped stars for the SMC are roughly consistent with other population synthesis studies \citep[e.g., ][]{hovis_2025}, but for the LMC we predict factor of $\sim2$ less. \cite{hovis_2025} assumed a constant star formation rate for each Cloud, which likely overestimates the star formation rate compared to the temporally and spatially resolved star formation history that we use in our population synthesis.

For each stripped star in our synthetic population, we assign physical parameters (e.g., component masses and orbital period) and synthetic photometry from the starburst population. We estimate optical photometric uncertainties by interpolating from the mean MCPS photometric uncertainties at the sources' apparent magnitudes. We do not assign UV photometric uncertainties, since we self-consistently compute them from the \textit{Swift} images after injecting the synthetic sources (see Section \ref{sec: UV photometry}.)

\subsubsection{Intrinsic UV excess of the synthetic population}\label{sec: intrinsic UV excess}
\begin{figure}
    \centering
    \includegraphics[width=\linewidth]{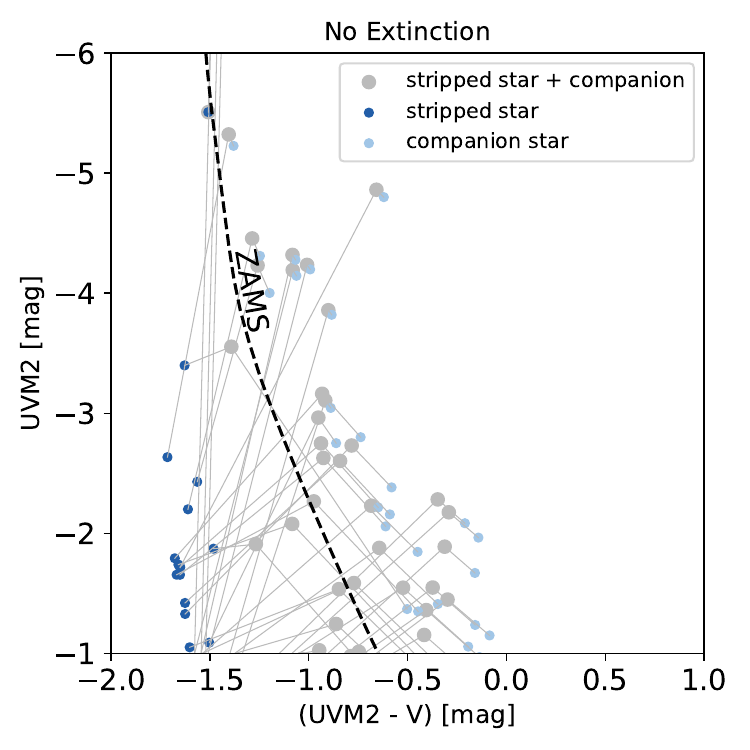}
    \caption{UV-optical CMD for a subset of synthetic stripped stars in the SMC. Dark blue markers indicate the stripped stars, light blue markers indicate the companions, and gray markers indicate the unresolved sources. The gray line connects each stripped star to its companion and unresolved source. The dashed line indicates the theoretical ZAMS. About 1/3 of the unresolved sources falls bluewards of the ZAMS and thus are UV-excess sources.}
    \label{fig: mock pop cmd}
\end{figure}

Figure \ref{fig: mock pop cmd} shows an example UV-optical CMD without extinction for a subset of synthetic sources in the SMC. The stripped stars lie bluewards of the theoretical ZAMS and exhibit UV excess, while the companions lie redward of the ZAMS. The combined unresolved sources fall on both sides of the ZAMS, depending on the relative UV and optical contributions from the stripped star and companion. 

We find that roughly 1/3 of the total synthetic stripped-star binaries exhibit intrinsic UV excess. That is, of the 3,505 (1,019 in the SMC and 2,486 in the LMC) simulated systems, 1,416 (336 in the SMC and 1080 in the LMC) have synthetic colors that are bluer than the theoretical ZAMS in 4 or more UV-optical color combinations. This is consistent with the predictions from \citetalias{drout_2023}, who estimated that $\sim26\,\%$ of stripped-star binaries are expected to show UV excess when assuming that half of binaries survive CEE; this value drops to $\sim 11\,\%$ when only 10\% of binaries are assumed to survive CEE.

The intrinsic UV-excess population is made up of 1,141 (265 in the SMC and 876 in the LMC) stripped stars hosting MS companions, 0 stripped stars hosting post-MS companions, and 275 (71 in the SMC and 204 in the LMC) stripped stars hosting compact object companions. In other words, $25-40\,\%$, $0\,\%$, and $95-100\,\%$ of the stripped stars with MS, post-MS, and compact object companions  show intrinsic UV excess. $\sim 5\,\%$ of stripped stars with compact object companions do not show UV excess because they are in the contraction or expansion phase. During this phase, the stripped stars are puffed up and have lower effective temperatures ($T_{\rm eff} < 30,000$ K) and larger radii ($R_{\rm stripped} \gtrsim 3-5 \,R_\odot$). This phase can last up to $\sim10 \,\%$ of a stripped star's lifetime \citep[e.g., ][]{Gotberg_2018, dutta_2024}. As shown in S2 (Figure S8) of \citetalias{drout_2023}, for stripped star + MS systems that show intrinsic UV excess, stripped stars more massive than $> 2\,M_\odot$ typically host MS companions with masses $M_{\rm MS} \lesssim 2 M_{\rm stripped}$, and less massive ($\lesssim 1\,M_\odot$) stripped stars typically host companions with masses $M_{\rm MS} \lesssim 4 M_{\rm stripped}$.

Even in the most ideal case, roughly 2/3 of stripped star binaries in our model cannot be detected via a UV-excess survey, because they are not UV-excess sources. Whether a system has intrinsic UV excess depends heavily on the type and mass of the companion. We further explore this in Section \ref{sec: companion discussion}.

\subsubsection{Spatial distribution of stripped stars}\label{sec: sptial distribition}
\begin{figure*}
    \centering
    \includegraphics[width=\linewidth]{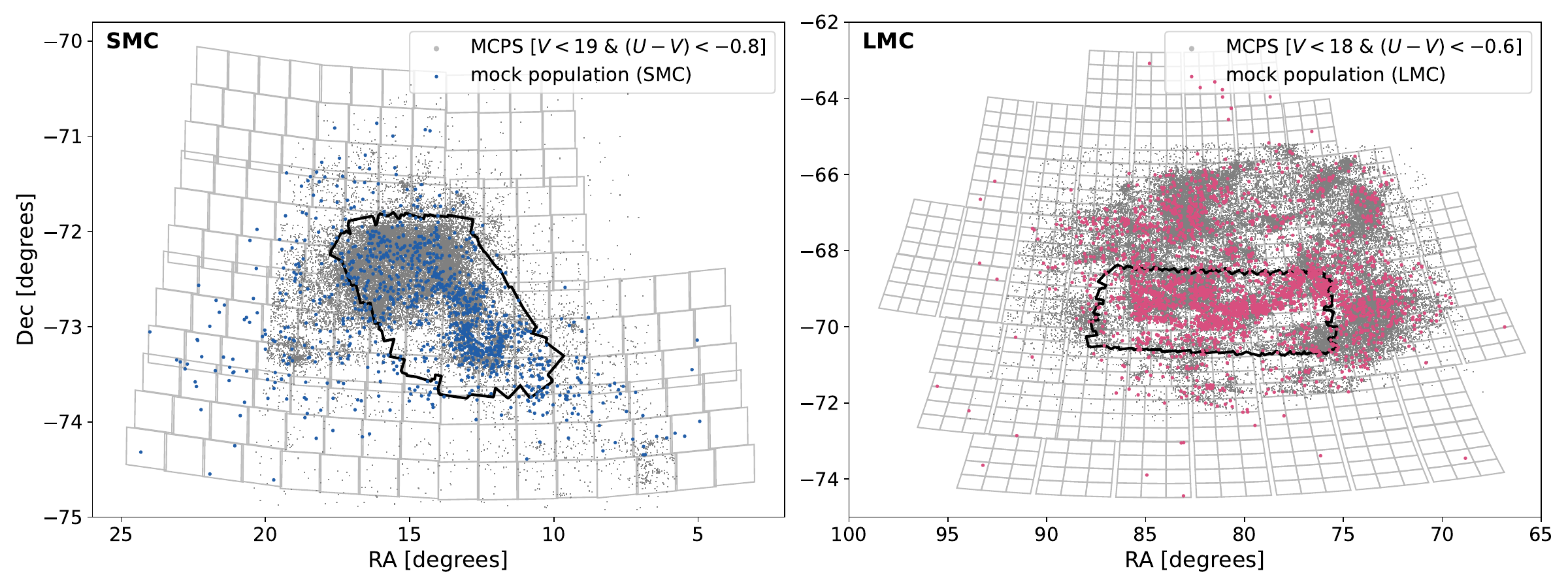}
    \caption{Spatial distribution of the synthetic stripped-star binaries in the SMC (left) and LMC (right). Each box represents a VMC subregion that \cite{rubele_2018} and \cite{Mazzi_2021} derived an average star formation history for. Gray markers indicate young MS stars from MCPS. Blue/pink markers indicate simulated stripped-star binaries. The black line shows the footprint of the SUMaC images used to perform UV photometry in SUMS.}
    \label{fig: mock stripped population spatial distribution}
\end{figure*}

We spatially distribute the simulated stripped stars following the 2D distribution of young-massive stars observed by MCPS. We identify young-massive stars as those with $V < 19$ mag and $(U-V) < -0.8$ mag (Vega) for the SMC, and  $V < 18$ mag and $(U-V) < -0.6$ mag for the LMC. These cuts select MS stars with masses $\gtrsim 4\,M_\odot$; we adopt slightly different selections for the LMC and SMC because the LMC contains more massive stars and has higher extinction on average. Within each VMC survey subregion, we draw RA and Dec coordinates for the simulated stripped stars from a smoothed 2D spatial distribution of young MS stars, such that stripped stars are more likely to fall in regions where other young stars are concentrated. Figure \ref{fig: mock stripped population spatial distribution} shows the resulting spatial distribution of the simulated stripped stars in the SMC and LMC. The stripped stars are distributed similarly to the young MS stars, with some differences (e.g., the distribution of sources across different regions) that arise from stripped stars typically being younger.

We assign distances to each stripped star by drawing from an exponential distribution characterized by a scale height $h_s$. That is, stellar distances along the line-of-sight are drawn from a distribution $p(z) \propto e^{-|z|/h_s}$, where $z$ is defined relative to the mean distance to each galaxy. The total distance to a source is $d = D + z$, where $D$ is the average distance to each Cloud. We take $D = 60.6$ kpc for the SMC \citep{hilditch_2005} and $D = 50$ kpc for the LMC \citep{Pietrzyski_2013}.

We adopt $h_s = 1$ kpc for the SMC and $h_s  = 0.6$ kpc for the LMC \citep[e.g.][]{subramanian_2009}. Both the SMC and LMC have irregular stellar morphologies as a result of their shallow gravitational potentials, recent encounters between the two galaxies, and ongoing gravitational interactions with the MW. Our modeling should not be interpreted as an assumption that stripped stars are distributed in an exponential disk. Rather, it serves to distribute stripped stars along the line of sight in a manner that is approximately consistent with the galaxies' vertical extent. Although this extent is small compared to the distance to the Clouds, the distribution of stripped stars along the line of sight is important because it determines what fraction of the total extinction column is in front of each source. For the LMC, we model a tilted stellar disk, with an inclination of $i = 35$ degrees (where $i = 0$ degrees corresponds to a face-on disk), and an line of node position angle of $\Theta = 122$ degrees (see \citealt{van_2001} for further details). The orientation of the SMC is not well constrained in the literature, and thus we model it as a face-on disk; see Appendix \ref{apdx: extinction calculation} for further discussion.

\subsection{Extinction}\label{sec: extintion}
Extinction is an important component of the forward model, because stripped stars with significant foreground extinction will appear redder than the ZAMS. When selecting photometric candidates, \citetalias{ludwig_2025} do not measure extinction on a star-by-star basis. Instead, all stars are de-reddened using an average extinction of $A_V = 0.38$ mag (LMC) and $A_V = 0.22$ mag (SMC). This will overestimate extinction for some sources that are in front of the dust, and underestimate it for those behind it. Stripped stars with too underestimated correction will not be recovered by UV-excess searches since they will appear redder than the ZAMS, whereas MS stars with overestimated extinction will appear bluer, possibly contaminating the sample of stripped-star candidates (see discussion in Section 5.9 of \citetalias{ludwig_2025}).

Below, we describe our treatment of extinction in the forward model. We assume that all extinction is due to intervening dust along the line of sight, neglecting any extinction from circumstellar material.

\subsubsection{Extinction curves}\label{sec: extinction curve}
\begin{figure}
    \centering
    \includegraphics[width=\linewidth]{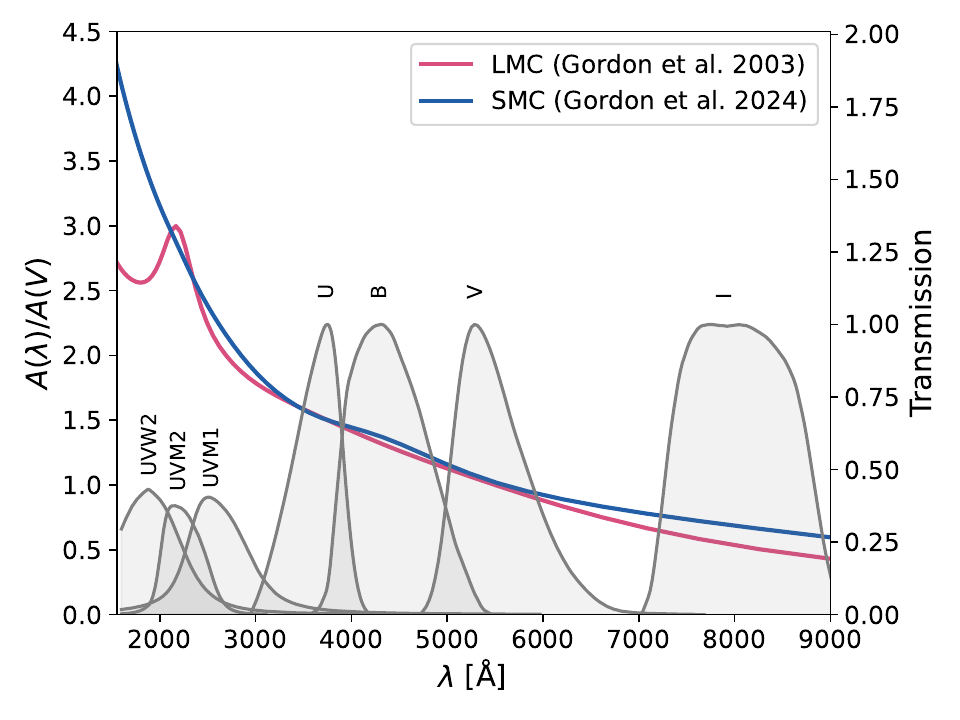}
    \caption{Average extinction curves for the SMC \citep{Gordon_2024} and LMC \citep{Gordon_2003}. The gray lines show the transmission curves for the optical ($U$, $B$, $V$, $I$) and UV ($UVW2$, $UVM2$, $UVW1$) bandpasses.}
    \label{fig: extinction curve}
\end{figure}

We adopt average extinction curves derived for the SMC by \cite{Gordon_2024} and LMC by \cite{Gordon_2003}; these are shown in Figure \ref{fig: extinction curve}. Using filter transmission curves \citep{rodrigo_2012, rodrigo_2020, rodrigo_2024}, we compute $R_\lambda = A_\lambda /E(B-V)$ (Table \ref{table: extinction values}). We compute $E(B-V)$ by integrating the spectrum of a star \citep{lejeune_1998}, and $A(\lambda)$ by evaluating the average SMC and LMC extinction curve for each UV and optical filters.
\begin{table}[h!]
\centering
\begin{tabular}{ c c c } 
 \hline
Photometric band & $R_\lambda$ (SMC)& $R_\lambda$ (LMC)  \\ \hline
$UVW2$ & 9.522& 8.737\\
$UVM2$ & 8.414 & 9.091\\
$UVW1$ & 6.774 & 6.465\\
$U$ & 4.674 & 4.835 \\
$B$ & 4.168  & 4.125 \\
$V$ & 3.120 & 3.148\\
$I$ & 2.129 & 1.731 \\
 \hline
\end{tabular}
\caption{$R_\lambda$ of relevant UV and optical bandpasses for the SMC and LMC. We use the average extinction curves for each Cloud \citep{Gordon_2003, Gordon_2024} in our calculations.}
\label{table: extinction values}
\end{table}

\subsubsection{Deprojecting the 2D dust map}\label{sec: 3d dust model}
\begin{figure*}
    \centering
    \includegraphics[width=1\linewidth]{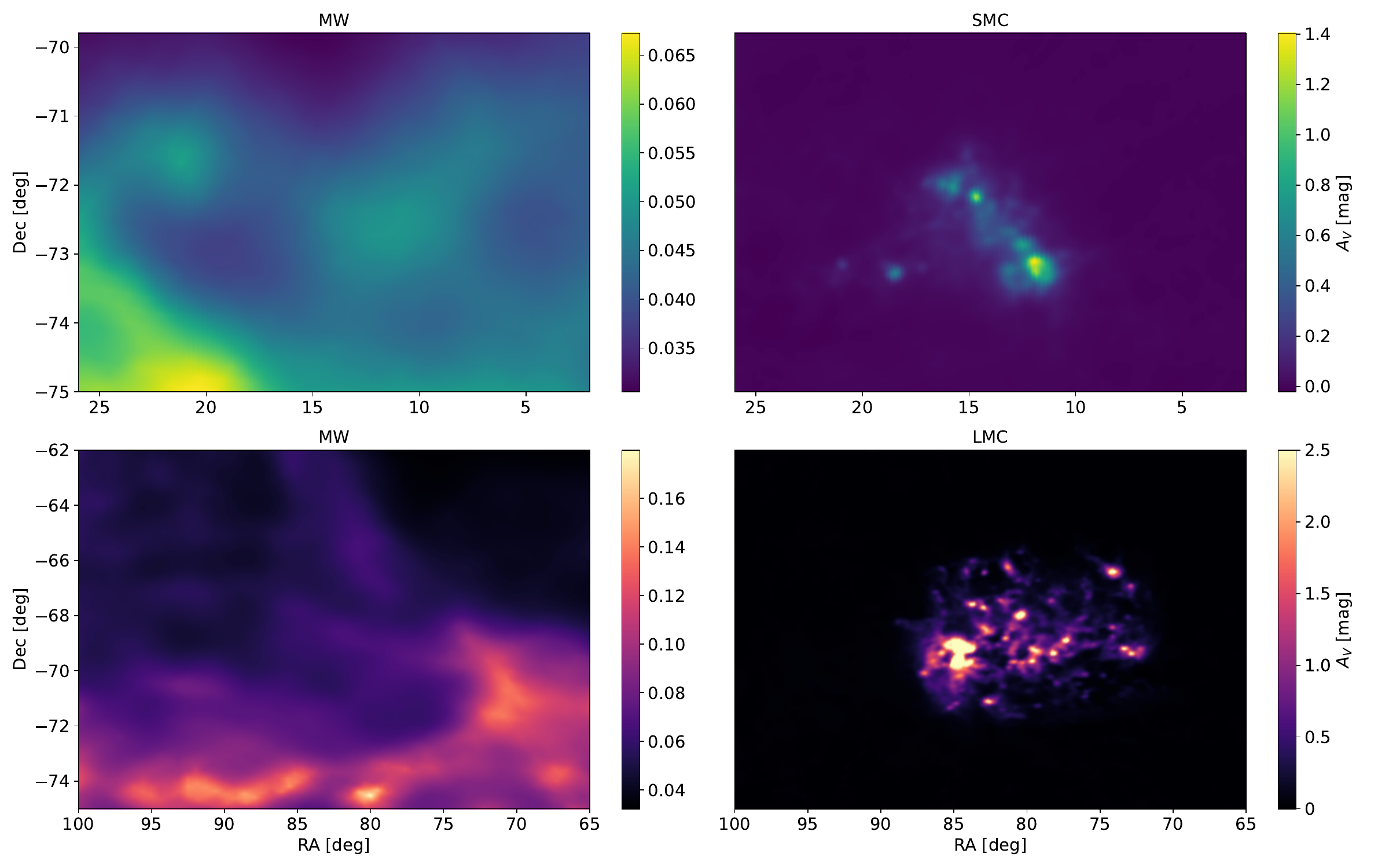}
    \caption{Foreground (left) and Magellanic (right) 2D extinction maps for the SMC (top) and LMC (bottom). These maps are computed by integrating the 3D dust maps from \cite{edenhofer_2024} out to 2 kpc (representing the Galactic foreground; left) and subtracting this from the \cite{sfd_dust} dust map (representing the total extinction along the line of sight). The right panels show the total predicted extinction within the MCs.}
    \label{fig: dustmaps}
\end{figure*}

The 3D structure of dust within the SMC and LMC is uncertain. However, the total integrated extinction column is well constrained by dust emission measurements. We take the extinction map computed by \citeauthor{sfd_dust} (\citeyear{sfd_dust}; hereafter SFD) as the total extinction to infinity. This contains both a foreground component due to dust in the MW, and a component representing the total column within the SMC and LMC. We take the dust map from \cite{edenhofer_2024} and integrate it out to 2 kpc to obtain the foreground contributions from the MW. Assuming that the majority of the dust along the line of sight is in the MW or the MCs -- with negligible dust in between or behind -- we can subtract the MW foreground from the SFD map to obtain an integrated 2D extinction map of the MCs:
\begin{equation}\label{eq: 2d dust}
    E(B-V)_{\rm{MCs}} = E(B-V)_{\rm{SFD}} - E(B-V)_{\rm{MW}}
\end{equation}
where $E(B-V)_{\rm{SFD}}$ is the SFD reddening and $E(B-V)_{\rm{MW}}$ is the MW foreground, evaluated from the \cite{edenhofer_2024} 3D dust map integrated out to $d = 2$ kpc. Figure \ref{fig: dustmaps} shows the MW dust map (i.e., foreground) and the SMC and LMC 2D dust maps. The total foreground dust contribution is relatively smooth compared to the dust within the MCs, which is more patchy. The foreground dust contribution is small and contributes $\lesssim 10\,\%$ to the total extinction along most lines of sight, but it is important because it affects all sources in the MCs. The LMC has much more dust compared to the SMC, with the maximum color excess of 2.5 mag, compared to the maximum of 1.4 mag in the SMC.
The total extinction values shown in Figure~\ref{fig: dustmaps} are larger than the values inferred by studies that estimated extinctions from the colors of hot stars  \citep[e.g., ][]{harris_2004, harris_2009, hagen_2016}. However, this is expected because the extinction values measured in those studies are not representative of the total extinction along the line of sight, since most of the stars are not behind the full dust column.

With the total Magellanic extinction at a given sky position constrained by Equation \ref{eq: 2d dust}, we assume that the dust is distributed along the line of sight exponentially with a scale height $h_d$. Along each line of sight, we require that the integrated total extinction within the MCs equal extinction calculated by subtracting the foreground from the SFD map. To obtain a extinction value for a source at distance $z$ (along the line of sight) from the center of the MCs, we integrate the Magellanic extinction in front of the source and add the foreground contribution from the MW:
\begin{equation}\label{eq: extinction}
    E(B-V)(z) = E(B-V)_{\rm MW}+ \int_{-\infty}^{z} E(B-V)_{z'} \, dz'.
\end{equation}
Here, $E(B-V)_z \propto e^{-|z|/h_d}$ is the dust density along the line of sight, normalized such that it integrates to the value from Equation \ref{eq: 2d dust}.

We assume $h_d = 0.3$ kpc for both the SMC and LMC. We chose this value after comparing predicted optical CMDs simulated with a range of $h_d$, to observed CMDs. As explored in Appendix \ref{apdx: extinction calculation}, the width of the MS is diagnostic of $h_d$, and we find $h_d = 0.3$ kpc best reproduces the MCPS data in the upper MS. Like the stars, dust in the SMC and LMC has irregular morphology and are not distributed in a thin disk. Our results are nevertheless only weakly sensitive to the choice of $h_d$, since the total integrated extinction at each spatial position is constrained by the empirical 2D dust map.

Using Equation \ref{eq: extinction}, we compute $E(B-V)$ for each simulated source based on its position and distance along the line of sight. Using $R_\lambda$ (Table \ref{table: extinction values}), we compute the extinction $A_\lambda$ for each bandpass:
\begin{equation}\label{eq: reddening relation}
    A_\lambda = R_\lambda E(B-V).
\end{equation}
We compute apparent magnitudes for each source using its distance and add the effects of extinction:
\begin{equation}\label{eq: magnitude}
    m = M + 2.5\log\left[\left(\frac{d}{10}\right)^2\right] + A_\lambda,
\end{equation}
where $m$ is the apparent magnitude of the simulated source, and $M$ is the intrinsic absolute magnitude assigned to each source in Section \ref{sec: synthetic photometry and final outputs}.

\subsection{Mock photometry}\label{sec: UV photometry}
We only simulate the UV photometry using mock observations. For the optical, we take apparent magnitudes computed by Equation \ref{eq: magnitude} as the mock observed optical photometry. We assume that every source with well-measured UV photometry will also have accurate optical photometry, because the \textit{Swift}-UVOT PSF (FWHM $\sim 2.5''$) is significantly larger than the MCPS PSF ($\sim 1''$). We thus expect the photometry from \textit{Swift}, which is more strongly affected by crowding than the optical, to set the sensitivity of the UV-excess sample. We note, however, that \citetalias{ludwig_2025} point out some quality issues in some of the MCPS photometry, including (i) individual photometric bands which are inconsistent with stellar SEDs, (ii) sources with anomalously faint I-band magnitudes, and (iii) missing optical coverage caused by gaps in the MCPS scanning pattern. Our forward model does not attempt to model these effects.

To forward model the UV photometry of SUMS, we inject simulated stripped-star binaries into the \textit{Swift}-UVOT images and perform mock photometry using the photometric methods described in \citetalias{ludwig_2025}.
 
\subsubsection{Source injection}\label{sec: source injection}
We convert the magnitude of each synthetic stripped-star binary into a detector count rate and model its flux profile using the \textit{Swift}-UVOT PSF. We then inject each synthetic source into level 2 \textit{Swift}-UVOT images for all three UV bandpasses. As noted by \citetalias{ludwig_2025}, a number of the images obtained by the SUMaC survey have tracking issues that distort the PSF. Rather than attempting to model these distortions, we remove these images from our analysis following \citetalias{ludwig_2025}. In total, we remove 82 images from the SMC (22 in $UVM2$, 27 in $UVW1$, and 33 in $UVW2$) and 154 from the LMC (46 in $UVM2$, 33 in $UVW1$, and 55 in $UVW2$). Additionally, we mask regions of the images, where extremely bright sources saturate the detector, making it difficult to perform photometry. We use the same masking technique as described in \citetalias{ludwig_2025}, and mask a 12'' region around sources with count rates of $\sim 65$ counts/s or higher within a 5'' circular aperture centered on a given source. We mask less than $\sim 1\,\%$ of the SUMS footprint.

Even after masking and removing problematic images, most sources lie in regions that were observed more than once by \textit{Swift}; either due to (i) multiple exposures taken in a single visit, (ii) survey revisiting the same region in different epochs, or (iii) sources falling in overlapping regions between survey fields. Instead of stacking the images before performing photometry, we repeat the injection and recovery process for every image separately and average the results. We use this approach to keep our photometry consistent with \citetalias{ludwig_2025}.

Since the flux calibration pipeline (HESARC; \cite{heasarc}) and CALDB \citep{swift_caldb} work in Vega magnitudes, we first convert the reddened synthetic magnitudes from AB to Vega using the conversions in Table \ref{table: AB vega conversion}. 
\begin{table}[h!]
\centering
\begin{tabular}{ c | c c } 
 \hline
photometric band & AB - Vega & ZPT \\ \hline
$UVW2$ & 1.705& 17.38\\
$UVM2$ & 1.659  & 16.85\\
$UVW1$ & 1.489 & 17.44\\
 \hline
\end{tabular}
\caption{AB to Vega magnitude conversion \citepalias{ludwig_2025} and the zero-point offset (ZPT) for each \textit{Swift}-UVOT filter \citep{breeveld_2010}.}
\label{table: AB vega conversion}
\end{table}
We then convert the Vega magnitudes into count rates using:
\begin{equation}\label{eq: mag to count}
    C = 10^{\frac{\rm{ZPT} - \textit{m}}{2.5}},
\end{equation}
where $C$ is the count rate in counts/s, ZPT is the zero point (Table \ref{table: AB vega conversion}), and $m$ is the apparent magnitude in Vega.

Equation \ref{eq: mag to count} gives the count rate for each UVOT filter assuming a perfect detector (i.e., $100\,\%$ detector efficiency and no coincidence loss). In practice, we must correct the input count rates for detector inefficiencies. This correction depends on factors such as detector age at the time of observation, the filter, the flux of the source, and the region of the detector which the source falls on. We use the \textit{Swift}-UVOT HEASARC calibration tools, following the process described in \citetalias{drout_2023} and \citetalias{ludwig_2025} to model these corrections for each source in each image it is injected into.  
We use the \texttt{uvotcoincidence} function, which corrects for coincidence losses, \texttt{uvotlss}, which computes large scale sensitivity corrections, and \texttt{uvotflux}, which converts detector count rates to Vega magnitudes.

To determine the appropriate input flux, we compute an initial guess for the corrected count rate using Equation \ref{eq: mag to count}. We then pass it through the calibration pipeline and iteratively adjust the input count rate until the calibrated magnitude matches the desired value within 0.005 mag. This approach ensures that deviations between the injected and recovered UV photometry are purely a result of crowding and source confusion, not systematic errors in flux calibration. We thus implicitly assume that the UVOT flux calibration is accurate.

\begin{figure*}
    \centering
    \includegraphics[width=0.9\linewidth]{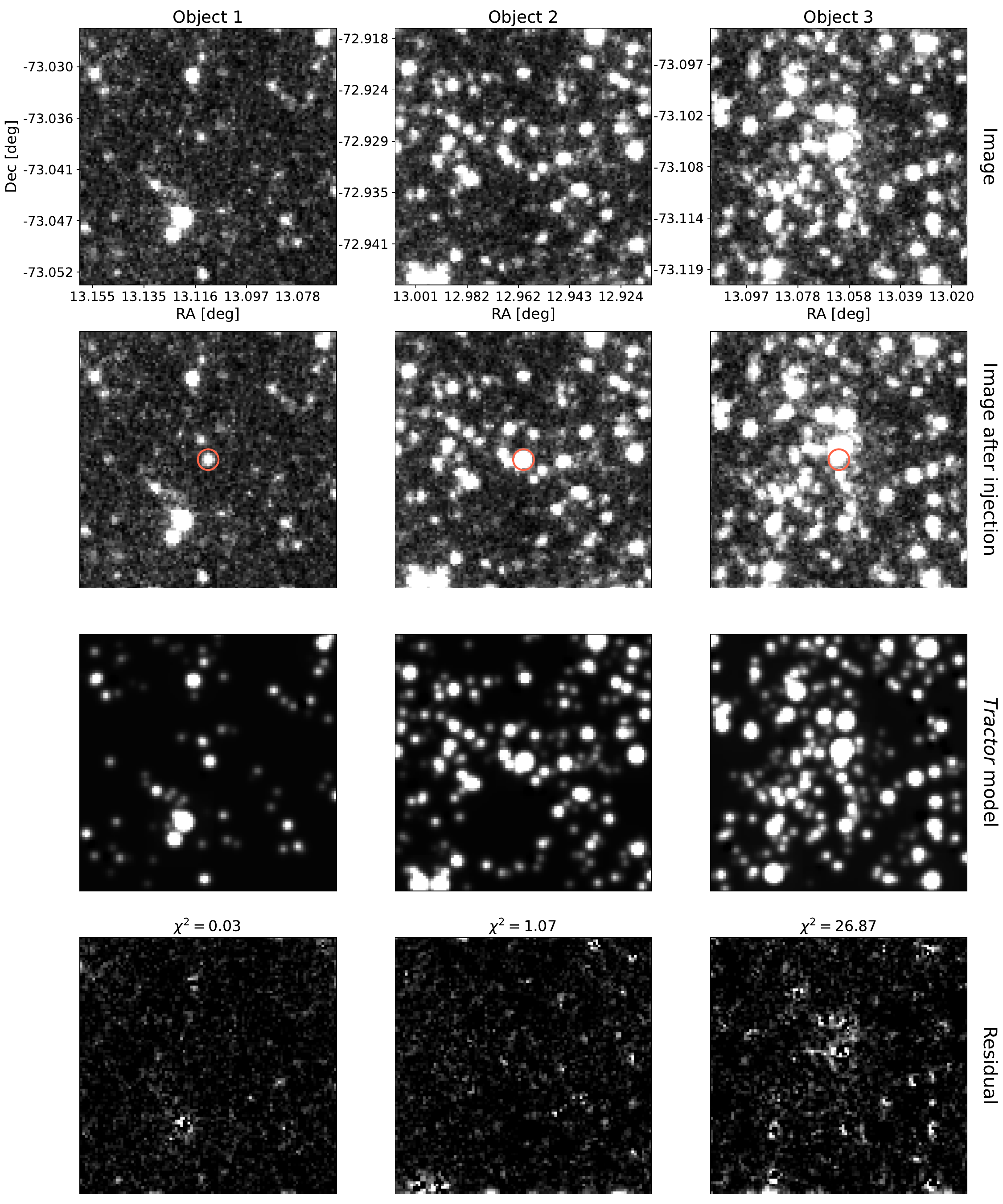}
    \caption{Top: cutouts of three \textit{Swift}-UVOT images centered on coordinates where we inject the synthetic sources, highlighting cases of low, moderate, and high source density. Second row: same set of images after the synthetic source -- marked with an orange circle -- is injected. Third row: models of the injected images produced by \textit{the Tractor}. Bottom: residual images, computed by subtracting the model (third row) from the data (second row). For each injected source, we compute $\chi^2 = \frac{|f_{\rm input} - f_{\rm output}|^2} {f_{\rm err}^2}$, where $f_{\rm input}$, $f_{\rm output}$, and $f_{\rm err}$ are the input flux, output flux, and flux error.    
    \textit{The Tractor} performs well in regions of low source density (left), moderately well in the moderate-density fields (middle) and poorly in the high-density fields (right). } 
    \label{fig: tractor exmaples}
\end{figure*}

To determine the shape of each source, we use the 2D pixelated PSF from \citetalias{ludwig_2025}, which models the \textit{Swift}-UVOT PSF following \cite{breeveld_2010}. The PSF for each bandpass is modeled separately using an analytic two-component model: the central 5 arcsec is represented by a Moffat profile and the outer 25 arcsec is represented by a double Gaussian profile. We assume the PSF is constant across all images for a given bandpass.  We inject each simulated source by adding the model profile to the science images in each bandpass.  This PSF model used to create the synthetic sources is the same model employed for \textit{The Tractor} photometry. This makes injected sources more symmetric and idealized compared to some of the real sources, which may have distorted PSFs. See the second row of Figure \ref{fig: tractor exmaples} for examples of injected sources.

\subsubsection{Performing mock photometry on injected images}\label{sec: mock photometry on images}
Using the photometric pipeline from \citetalias{ludwig_2025}, we measure the UV count rates for simulated injected sources using \textit{the Tractor}\footnote{https://thetractor.readthedocs.io/en/latest/intro.html} \citep{Lang_2016}. \textit{The Tractor} is a forward-modeling package that performs forced PSF photometry by modeling entire astronomical images. Unlike traditional aperture photometry methods, \textit{the Tractor} models all sources within an image simultaneously, making it more effective at de-blending crowded sources. This makes \textit{the Tractor} an optimal tool for performing UV photometry in the MCs, where typical separations between sources are smaller than the $2.5$ arcsec FWHM of the \textit{Swift}-UVOT PSF. 

We use \textit{the Tractor} in its forced photometry mode, which requires a set of input coordinates, where we expect sources to exist. Following \citetalias{ludwig_2025}, we provide the coordinates of optical sources from the MCPS catalog with $U< 20.5$ mag (or $B < 20.5$ mag when $U$ band photometry is not available). In addition, we also provide the coordinate of the synthetic stripped-star binary that we injected into the image. To account for astrometric uncertainties between the MCPS source coordinates and WCS of the \textit{Swift}-UVOT images, the \citetalias{ludwig_2025} pipeline allows the source coordinates in \textit{the Tractor} to vary within a 2D Gaussian probability distribution with a standard deviation of $\sigma = 0.05''$ centered around the input coordinates we provide.

To reduce runtime, we only model a small $100 \times 100$ pixel cutout for each image, centered on the injected source. Figure \ref{fig: tractor exmaples} shows several representative examples of the injected images and the corresponding models from \textit{the Tractor}.
The photometry is most reliable for moderately bright sources in non-crowded regions and least reliable for dim sources in crowded regions as predicted by \citetalias{ludwig_2025} (Section 3.4).

Once we have modeled each image using \textit{the Tractor}, we convert the output fluxes into magnitudes using the standard \textit{Swift}-UVOT HEASARC calibration tools. Following \citetalias{drout_2023} and \citetalias{ludwig_2025}, we use the \texttt{uvotcoincidence}, \texttt{uvotlss}, and \texttt{uvotflux} functions. These functions compute several quality flags, which identify sources that are saturated (`SATURATED'), fall in a region with abnormally low count rate (`SSS'), or fall within 5'' from the edge of the detector (`EDGE'). Such sources often lie in problematic regions of the detectors and are discarded.

At this stage, we follow \citetalias{ludwig_2025} and add an additional $5\,\%$ systematic error to the magnitudes to account for variations in the \textit{Swift}-UVOT PSF across images. We perform the same quality cuts on the resulting photometry as described in \citetalias{ludwig_2025}. We discard sources with magnitude uncertainties greater than 0.36 mag. We calculate the ``residual fraction'' (RF) as defined in \citetalias{ludwig_2025}. This is the residual (i.e., the difference between the flux from \textit{the Tractor} model and the \textit{Swift} image) divided by the total count rate assigned to a given source. We require  $\rm RF < 0.3$. We also remove sources which deviated from their input coordinates by more than 1 arcsec, which indicates a possible mismatch between the optical and UV source.

\subsubsection{Combining multiple measurements of a single source}\label{sec: combining mulitple measurements}
Following \citetalias{ludwig_2025}, for each source, we average the measurements from all of the images for a given filter. To perform outlier rejection, we compute the modified $Z$-score \citep{Hoaglin_2013}:
\begin{equation}
    Z_i = 0.6745 \frac{|m_i - m_{\rm median}|}{\rm{mad}},
\end{equation}
where $Z_i$ is the modified $Z$-score for the $i$-th image in a given filter for a given source, $m_i$ is \textit{the Tractor} magnitude for the $i$-th image, $m_{\rm median}$ is the median magnitude of all observations of a source for a given filter, and $mad$ is the median absolute deviation, defined as $\rm{median}\{|m_i - m_{\rm median}|\}$. We reject outliers by removing measurements with $Z_i > 3.5$, and we compute the weighted average of all remaining images. We define the weights as the inverse quadrature sum of the residual fraction and the uncertainty of the magnitudes:
\begin{equation}
    m_{\rm avg} = \frac{1}{n}\sum_{i = 0}^{i=n} {\frac{m_i}{\sqrt{RF_i^2 + m_i^2}}},
\end{equation}
where $n$ is the total number of observations we average over. We compute uncertainties on the magnitudes using the standard uncertainty propagation for weighted averages. Finally, we convert the Vega magnitudes into AB magnitudes using the conversions from Table \ref{table: AB vega conversion}. We repeat this for each filter for all simulated sources. 

\begin{figure}
    \centering
    \includegraphics[width=1\linewidth]{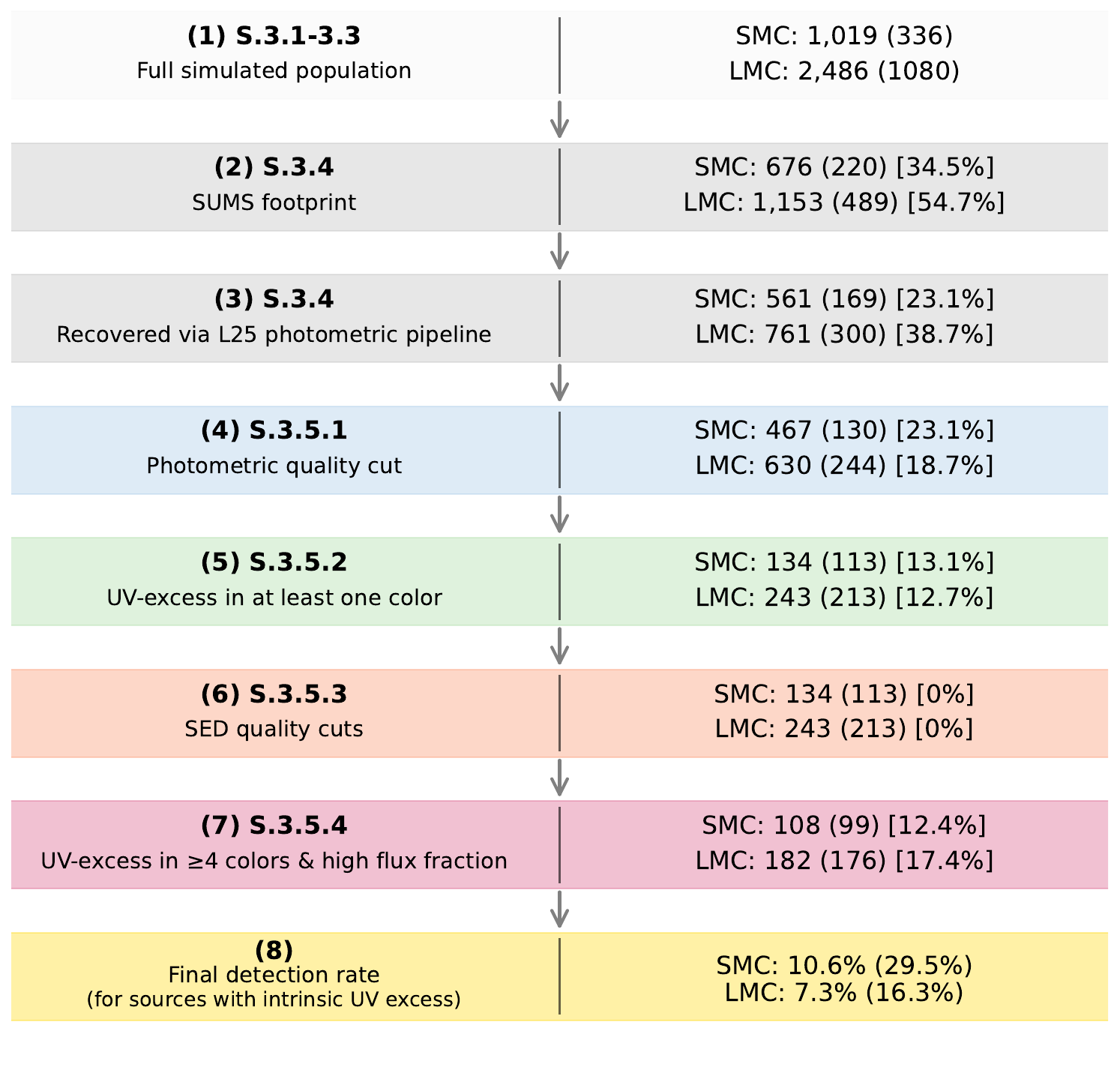}
    \caption{Flowchart of the forward model. Each box represents a stage in the forward model of the SUMS selection criteria \citepalias{ludwig_2025} where sources are lost. The values represent the number of simulated stripped stars that remain after each cut. The values in parentheses indicate sources that have intrinsic UV excess in four or more UV-optical colors. The values in brackets indicate the fraction of intrinsic UV-excess sources rejected at each step relative to the previous one. Box (1) represents the full population and shows the total number of stripped-star binaries predicted in the MCs. Boxes (2) and (3) represent the injection and recovery process. Boxes (3) through (7) represent the quality cuts and UV-excess selection used to select candidates. In box (8), we quote the detection rate of the synthetic stripped star population when following the SUMS selection criteria. In the parenthesis, we quote the detection rate for systems when only considering systems which show intrinsic UV excess.}
    \label{fig: flowchart}
\end{figure}

\begin{figure*}
    \centering
    \includegraphics[width=\linewidth]{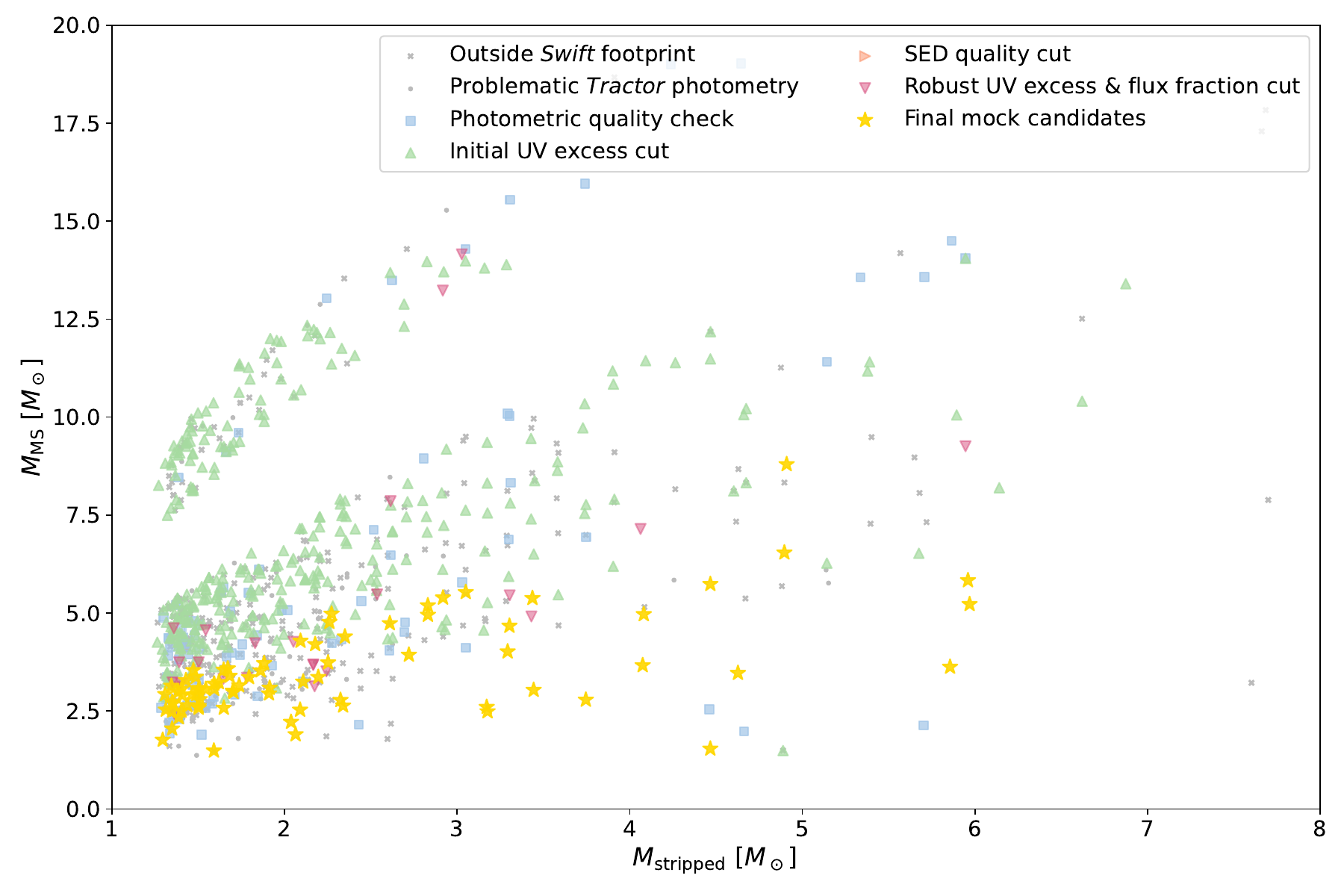}
    \caption{Masses of simulated stripped stars and their MS companions for the SMC, with sources marked by the first step where they are rejected in the candidate selection process. The colors of the markers, described below, correspond to cuts made in the forward model. This is also depicted in the flowchart shown in Figure \ref{fig: flowchart}. The gray points represent sources that are outside of the SUMaC footprint or do not have photometry recovered by \textit{the Tractor}. Blue squares represent sources that are rejected due to bad photometry (Section \ref{sec: photometric quality cuts}). These cuts affect all masses of stripped stars and companions roughly equally. Green triangles represent sources that do not show UV excess in any UV-optical color combination (Section \ref{sec: initial uv excess cut}); orange sideways triangle show sources that were lost in the SED quality cuts (Section \ref{sec: SED quality cuts}); pink downward facing triangles represent sources that were rejected either because they do not show UV excess in more than three color combinations or have low flux fractions (Section \ref{sec: final uv excess cut}). The yellow stars represent sources that pass all candidate selection criteria and enter the final mock candidate population. 
    }
    \label{fig: candidate cuts}
\end{figure*}

Of the 3,505 synthetic stripped stars (1,019 in the SMC and 2,486 in the LMC) within the VMC survey footprint, we recover UV photometry for 1,322 systems (561 in the SMC and 761 in the LMC). 1,676 of the simulated stripped stars (343 in the SMC and 1,333 in the LMC) are lost because they lie outside of the SUMS footprint. The remaining 507 systems (115 in the SMC and 392 in the LMC) are excluded due to the lack of reliable UV photometry in all three UV bands, primarily as a result of crowding and/or poor photometric quality; see Section \ref{sec: sources lost due to photometry} for further discussion. Figure \ref{fig: flowchart} depicts the forward model as a flowchart and lists the number of sources recovered after each subsequent cut. Figure \ref{fig: candidate cuts} shows the earliest stage a synthetic source is lost in the forward model as a function of its stripped star mass and companion mass. The mock photometry is represented in box (2) and (3) of the flow chart and by gray points in Figure \ref{fig: candidate cuts}.

\subsection{Candidate selection}\label{sec: candidate selection}
We apply a set of photometric and SED quality cuts to our synthetic stripped-star population and check for detectable UV excess. We follow the candidate selection criteria used by \citetalias{ludwig_2025}, which we describe below.

\subsubsection{Photometric quality cuts}\label{sec: photometric quality cuts}
We only model sources with extinction-corrected UV magnitudes of $14$ mag $< m < 19$ mag (AB). These limits were chosen by \citetalias{ludwig_2025} based on the predicted magnitudes for intermediate-mass stripped star binaries in the MCs. We apply a uniform extinction correction of $A_V = 0.22$ mag in the SMC and $A_V = 0.38$ mag in the LMC. After the initial cut, we require all UV photometry to be detected at least at 5$\sigma$, which corresponds to photometric errors of less than 0.217 mag. We require that each source has at least two UV photometric points that pass the above cuts. Of the 1,322 (561 in the SMC and 761 in the LMC) simulated systems with photometry recovered by \textit{the Tractor}, 1,097 (467 in the SMC and 630 in the LMC) pass the above photometric quality cuts (Figure \ref{fig: flowchart}, box 4).

\subsubsection{Initial UV excess cut}\label{sec: initial uv excess cut}
We compute the UV excess for each source in all nine UV-optical color combinations: ($UVW2$, $UVM2$, $UVW1$) - ($B$, $V$, $I$). We define UV excess as 
\begin{equation}
    \rm{UV \, excess} = (UV - optical)_{\rm obs} - (UV - optical)_{\rm ZAMS},
\end{equation}
where $\rm (UV - optical)_{\rm obs}$ is the observed color and $\rm (UV - optical)_{\rm ZAMS}$ is the UV-optical color of the theoretical ZAMS at the same apparent UV magnitude. If a source's measured UV excess is greater than the photometric uncertainty in the UV-optical color, we classify the source as a UV-excess source. In this initial cut, we require that at least one of the nine colors show UV excess. At this stage, 720 (333 in the SMC and 387 in the LMC) of the simulated stripped stars are rejected because they lack detectable UV excess (Figure \ref{fig: flowchart}, box 5).

\subsubsection{Spectral energy distribution (SED) quality cut}\label{sec: SED quality cuts}
All sources that pass the above cuts are subjected to further photometric quality checks. We test for three possible problems: UV-optical mismatch, optically-red SEDs, and poor quality photometry. The SED checks performed in \citetalias{ludwig_2025} attempt to flag sources with spurious UV excess caused by issues such as UV-optical mismatch and variable stars. Although we do not expect many of our synthetic sources to be rejected by these quality checks given that our population only contains stripped-star systems, we perform the SED quality cuts to keep consistent with the selection function from \citetalias{ludwig_2025}. We summarize each of the cuts below; see Section 5.4 of \citetalias{ludwig_2025} for a complete description.

\textit{UV-optical mismatch: }this flags sources where the UV and optical photometry may not come from the same physical source. First, for systems with flat SEDs, we check that the UV photometry is consistent with the optical photometry. If the standard deviation between all optical bands is less than 0.217 mag, and similarly if the standard deviation between all UV bands is less than 0.217 mag, we classify a source as having a flat SED. For sources with flat SEDs, we check that the mean optical photometry is within 0.217 mag of the mean UV photometry. Second, we check for large jumps between adjacent UV and optical photometric points. We compute the color between the reddest UV band and the bluest optical band available for a given source. We compare this UV-optical color against the expected color computed from a theoretical SED taken from a composite stripped star + MS model grid, computed in S5.1.2 of \citetalias{drout_2023}. We remove sources that have colors that differ from the expected color by more than 0.217 mag. We find that typically no simulated stripped stars are flagged by this cut, since our simulated sources have well-defined coordinates and optical counterparts.

\textit{Optically red SEDs: }\citetalias{ludwig_2025} found that some of their observed sources have SEDs that become progressively redder at longer wavelengths. This is not consistent with expectations for stripped-star binaries with MS companions. We remove sources from the candidate sample if their flux densities increase toward redder wavelengths in two or more adjacent set of bands. This cut flags sources with spurious optical flux which we do not have in our simulated population. Thus, as expected, typically all simulated stripped stars pass this quality cut, with no more than two sources begin flagged in any given mock population.

\textit{Photometric outliers: }we test for photometric outliers caused by bad photometry. We compute colors between all adjacent bands and compare observed colors to a range of colors computed from theoretical SEDs. If a source falls outside the expected range for a typical stripped-star binary by more than 0.217 mag, we flag the color as suspect. If more than three colors are flagged as suspect, we remove the source from the candidate list. If two adjacent colors are flagged, we remove the photometric point in the middle. If one (or two non-adjacent) color(s) is flagged, we remove all of the photometric points associated with the suspect color(s). Typically no stripped stars are flagged in this step. This is unsurprising because this cut was designed by \citetalias{ludwig_2025} primarily to identify sources with outliers in their MCPS optical photometry, and by construction our optical photometry has no such outliers.

\subsubsection{Flux fraction and final UV excess cut}\label{sec: final uv excess cut}
Our final cut removes sources that are highly blended with bright neighbors. Following \citetalias{ludwig_2025}, we compute a ``flux fraction'' for each source, which is define as the fraction of UV flux that a source contributes to the total flux within a 5'' circular aperture centered on that source. Like \citetalias{ludwig_2025}, we require all sources to have flux fraction greater than $10\,\%$. For those with a neighboring source within $2.5''$, we require a flux fraction greater than $25\,\%$. This cut was designed by \citetalias{ludwig_2025} to ensure that the UV excess is attributed to the flux from the source itself and not due to neighboring UV-bright sources. 

Finally, we require that each candidate stripped star exhibits UV excess in at least four out of nine UV-optical CMDs following the selection criteria of SUMS. Of the 377 (134 in the SMC and 243 in the LMC) remaining candidates, 87 (26 in the SMC and 61 in the LMC) are rejected in this final selection cut (Figure \ref{fig: flowchart}, box 7). Fewer than 10 sources in each galaxy are rejected due to low flux fractions, with the majority being rejected due to a lack of robust UV excess across several color combinations.

\subsubsection{Candidate classification}\label{sec: candidate classification}
Following \citetalias{ludwig_2025}, we sort all candidates into four classes based on their degree of UV excess and photometric and SED qualities. 

\textit{``Very Blue'' vs. ``Blue''}: we classify a source as ``Very Blue'' if the average UV excess across all good bands is greater than 0.4 mag and ``Blue'' if it is greater than zero but less than 0.4 mag. 

\textit{``Excellent'' vs. ``Good''}: these classifications separate sources based on their photometric and SED quality. We classify a source as ``Excellent'' if it passes all of the following quality checks: the source has (i) an average UV flux fraction greater than $40\,\%$, (ii) at least 6 out of the 7 photometric bands in the SED, and (iii) no more than one adjacent pair of photometric bands in the SED that is inconsistent with the theoretical colors. All other sources are considered ``Good.'' 

Table \ref{table: candidates} summarizes the number of synthetic sources in each class. 
\begin{table}
\centering
\begin{tabular}{ c | c c c c} 
 \hline
Galaxy & classification & observed  & simulated & intrinsic \\ \hline\hline
SMC & VB-E & 27 ($9\,\%$)  & 43 ($39\,\%$) & - \\
  & VB-G & 76 ($25\,\%$) & 4 ($5\,\%$) & -\\
& B-E & 91 ($31\,\%$) & 55 ($50\,\%$)& - \\
  & B-G & 104 ($35\,\%$) &6 ($6\,\%$)& -\\\cline{2-5}
 & total & 298 & 108 & 336\\
 \hline
LMC & VB-E & 78 ($15\,\%$) & 90 ($49\,\%$) & -\\
  & VB-G & 235 ($45\,\%$) & 5 ($3\,\%$) & -\\
& B-E & 91 ($17\,\%$) &  85 ($47\,\%$) & -\\
  & B-G & 118 ($23\,\%$) & 2 ($1\,\%$) & -\\\cline{2-5}
   & total & 522 & 182 & 1080\\
 \hline
\end{tabular}
\caption{Total number of stripped-star candidates for each Cloud. The ``observed'' column shows the number of candidates from the SUMS catalog. The ``simulated'' column shows the number of simulated candidates recovered by the SUMS selection criteria described in Section \ref{sec: candidate selection}. The ``intrinsic'' column represents the underlying population before accounting for observational errors, distances, or extinction, which show intrinsic UV excess in 4 or more of the 9 UV-optical colors (see Section \ref{sec: intrinsic UV excess}). The fraction of sources in each class compared to the respective totals are listed in the parentheses. The classification criteria are described in Section \ref{sec: candidate classification}. We repeat the simulation three times to assess the impact of shot noise. The results for the other two trials are summarized in Appendix \ref{apdx: multiple runs}.}
\label{table: candidates}
\end{table}
For both Clouds, we find that the number of simulated sources classified as ``VB-E'' is roughly consistent with the observed population from \citetalias{ludwig_2025}, while in the SMC, the number of simulated sources classified as ``B-E'' underestimates the observed population by a factor of two. This suggests that many of the ``B-E'' candidates may be false-positives. This is consistent with the assessment from \citetalias{ludwig_2025} which shows ``Blue'' candidates are more likely to be false-positives (due to e.g., MS stars with over-corrected reddening). We also note that \citetalias{drout_2023} identified several contaminants in their spectroscopic sample (S3.7), indicating that some level of contamination is expected in UV-excess selected populations. The simulation also significantly underpredicts the number of ``Good'' candidates for both strong and weak UV-excess sources. This is not unexpected, since our simulated population always has well-behaved photometry in all four optical filters, but 
this shows that in our simulations, stripped stars with low flux fractions are rare. This suggests that the observed population in the ``Good'' classes may contain false-positive sources. We further discuss the purity of the SUMS candidates and possible contamination from young MS stars in Section \ref{sec: putiry}.

\begin{figure*}
    \centering
    \includegraphics[width=\linewidth]{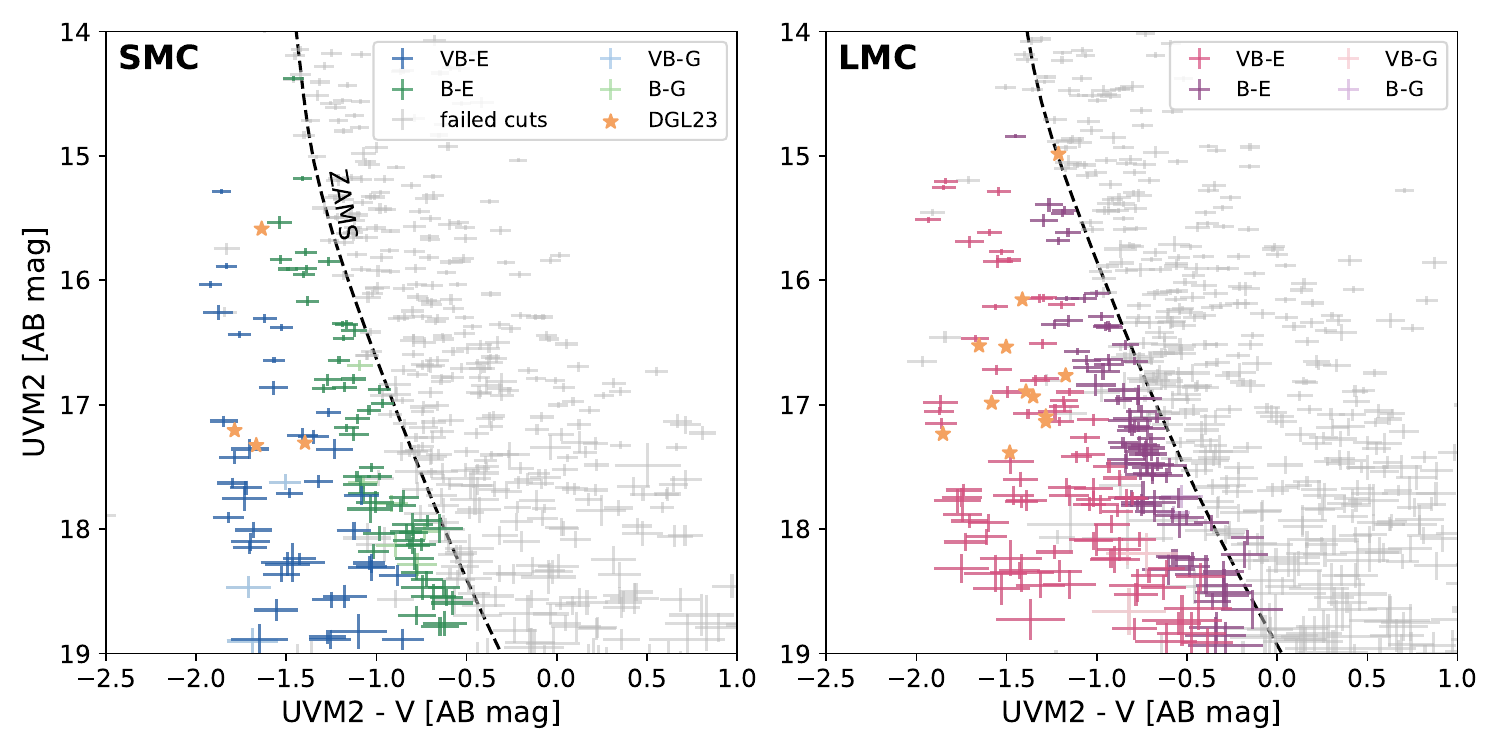}
    \caption{One of the nine UV-optical CMDs of the mock population for the SMC (left) and LMC (right). We correct the photometry by a uniform extinction of $A_V = 0.22$ mag for the SMC and $A_V = 0.38$ mag for the LMC following \citetalias{drout_2023} and \citetalias{ludwig_2025}. The dashed line indicates the theoretical ZAMS. Gray points show simulated stripped stars that pass photometric quality checks but do not make the final stripped-star catalog. The colored markers show simulated stripped stars that pass all candidate selection criteria, with the colors corresponding to its candidate classification. The orange stars show spectroscopically confirmed systems from \citetalias{drout_2023}.}
    \label{fig:mock_pop_cmd}
\end{figure*}

Figure \ref{fig:mock_pop_cmd} shows a CMD of the synthetic population. By construction, the ``Very Blue'' candidates exhibit larger UV excess and lie further away from the ZAMS than the ``Blue'' candidates. The ``Good'' candidates tend to be slightly fainter ($UV \gtrsim 16.5$) than the ``Excellent'' candidates, though there is no strong correlation between the CMD location and the distinction between ``Excellent'' and ``Good'' candidates. 
Of the sources that pass all quality cuts, $\sim90\,\%$ have intrinsic UV excess. The other $\sim10\,\%$ do not have intrinsic UV excess despite being true stripped-star binaries, and are only selected because crowding or overcorrected extinction result in overestimated UV fluxes.

We run the full simulation several times to assess the impact of shot noise due to e.g., the precise locations where we inject the simulated sources. We Monte Carlo sample a new underlying population of stripped stars for each run and assign spatial coordinates and distances for each source by drawing randomly from the respective probability distributions. The results are summarized in Appendix \ref{apdx: multiple runs}. For the remainder of the paper, we report the population statistics (e.g., number of stripped stars, detection rates) using the mean and standard deviation from the three iterations of the forward model.

\section{Mock-observed population}\label{sec: mock observed population}
After measuring mock photometry, applying quality cuts, and selecting sources with UV excess, we are left with $111\,\pm\,4$ (out of $1023\,\pm\,22$) simulated candidates in the SMC and $178\,\pm\,5$ (out of $2550\,\pm\,47$) in the LMC. This corresponds to a detection rate of $(10.9\,\pm\,0.5)\,\%$ for the SMC and $(7.0\,\pm\,0.2)\,\%$ for the LMC. Correcting for SUMS spatial coverage, this yields a detection rate of $(16.4\,\pm\,0.6)\,\%$ and $(14.8\,\pm\,0.6)\,\%$ for the SMC and LMC. Of the stripped stars that show intrinsic UV excess, we recover $(31.3\,\pm\,4.3)\,\%$ of sources in the SMC and $(15.3\,\pm\,0.6)\,\%$ in the LMC (or equivalently, $(47.2\,\pm\,3.8)\,\%$ and $(33.3\,\pm\,1.5)\,\%$ of simulated intrinsic UV-excess sources within the SUMS footprint).


\subsection{Companion flux contribution}\label{sec: companion discussion}
UV-excess searches are only sensitive to sources with detectable UV excess, and intrinsic UV excess of a stripped-star binary depends on the binary mass ratio and the nature of the companion. Here, we quantify how these factors effect the final detected population.

Of the $289$ ($111$ in SMC; $178$ in LMC) synthetic stripped stars which pass all selection criteria, $224\,\pm\,9$ ($85\,\pm\,8$ in SMC; $139\,\pm\,3$ in LMC) host MS companions and $65\,\pm\,6$ ($26\,\pm\,4$ in SMC; $39\,\pm\,4$ in LMC) host compact object companions. Given the distribution of companions in the original synthetic population ($\sim 88\,\%$ MS companions, $\sim 2\,\%$ post-MS companions, $\sim 10\,\%$ compact object companions), the companion-dependent detection rate of stripped stars is $(9.1\,\pm\,0.5)\,\%$ and $(6.0\,\pm\,0.2)\,\%$ for those hosting MS companions, $0\,\%$ for those hosting post-MS companions, and $(37.1\,\pm\,5.9)\,\%$ and $(18.5\,\pm\,2.2)\,\%$ for those hosting compact object companions, for the SMC and LMC, respectively. As predicted by \citetalias{drout_2023}, the UV-excess selection strongly favors stripped stars with compact object companions, because, unlike MS companions, they do not reduce the UV excess intrinsic to the stripped star. Consequently, while $\sim 88\,\%$ of the stripped stars in the underlying synthetic population host MS companions, only $\sim 77\,\%$ of the simulated sources recovered via the SUMS UV-excess criteria have MS companions. In contrast, while only $\sim 10\,\%$ of the underlying synthetic population have compact object companions, this fraction increases to $\sim 22\,\%$ for sources with detected UV excess.

\begin{figure*}
    \centering
    \includegraphics[width=\linewidth]{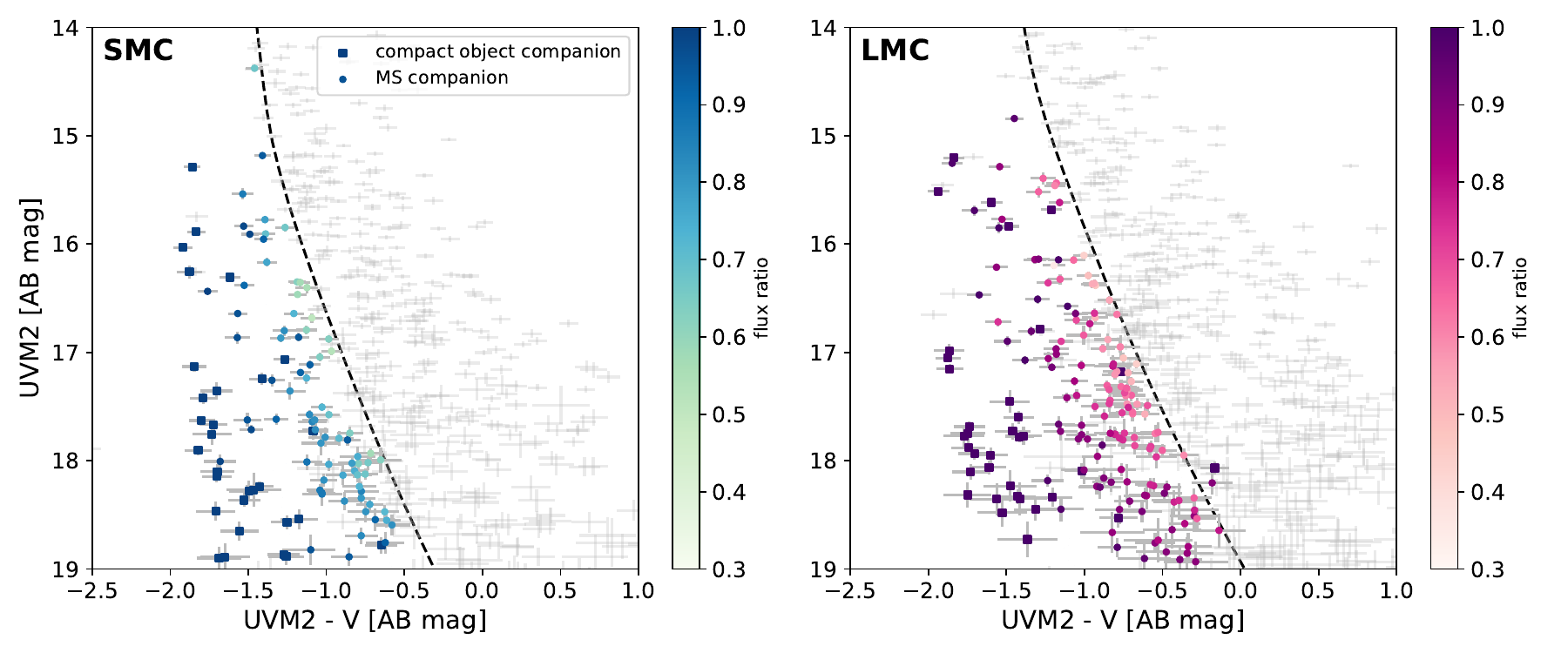}
    \caption{UV-optical CMD of the mock population for the SMC (left) and LMC (right). The photometry is de-reddened using a uniform extinction of $A_V = 0.22$ mag for the SMC and $A_V = 0.38$ mag for the LMC, following \citetalias{drout_2023} and \citetalias{ludwig_2025}. The gray markers show sources which do not pass the SUMS candidate selection criteria.
    The colored markers show sources in the final UV-excess sample, and the color of the markers correspond to the $UVM2$ flux ratio. The square markers indicate stripped stars with compact object companions and the circle markers indicate stripped stars with MS companions. }
    \label{fig: mock pop flux ratio}
\end{figure*}

We quantify the relative brightness of the stripped star to its companion using the ``UV flux ratio''--the fraction of the total UV flux that is contributed by the stripped star. Figure \ref{fig: mock pop flux ratio} shows one of the UV-optical CMDs with sources colored by their $UVM2$ flux ratios.
The candidates that lie further away from the ZAMS have higher flux ratios compared to the candidates near the ZAMS. Most of the synthetic stripped stars that meet the UV-excess criteria have photometry dominated by the stripped star (i.e., flux ratio $> 0.5$). This is roughly consistent with the predictions from \citetalias{drout_2023} who estimated that stripped stars must contributed $70-80\,\%$ of the total flux in the UV. The lower threshold measured from our forward model compared to previous predictions could be attributed to several factors: overestimated UV flux due to (i) crowding, and (ii) overcorrected reddening. Both of these factors are accounted for the first time in our study. These effects may cause sources with relatively low UV flux ratios ($0.5-0.7$) to appear bluer than they are intrinsically, allowing them to enter the final UV-excess sample even if they do not have true intrinsic UV excess (see Section \ref{sec: crowding}).

Similarly, we compute the ``optical flux ratio.'' We find that the optical flux ratio for our final detected sources range from $0.2 - 1$. This implies that a stripped-star binary can exhibit UV excess in a UV-optical CMD, even if the optical photometry is dominated by its companion. Some of these sources with low optical flux ratios but significant UV excess may be the ``class 3'' sources from \citetalias{drout_2023}.

The above flux ratios correspond to detectable mass limits for the stripped star and its companion, if we assume the limiting factor for detection is the UV flux ratio (i.e., no extinction, perfect photometry, non-crowded region). According to our new UV flux ratio threshold for UV-excess detection, stripped stars of $M_{\rm stripped} = 1.5 - 7\,M_\odot$ will be detectable if they host a MS companion with $M_{\rm MS} < 3.5 - 7\,M_\odot$ (see Section \ref{sec: mass /period distribution} for further discussion). Stripped stars with post-MS companions have UV flux ratio $> 0.5$ since the companion contributes negligible flux in the UV; however, these sources are red in the optical, which makes them redder than the ZAMS, making them more difficult to detect via UV excess. Stripped stars (not in the contraction or expansion phase) hosting compact object companions will always have UV flux ratio $\sim 1$ because the companion will contribute negligible light compared to the stripped star. Thus, they are always detectable via UV excess under ideal conditions. Stripped stars in their contraction or expansion phase, and thus appear puffed up, will not be detected in a UV-excess survey regardless of it companion type because the stripped star is faint in the UV due to its larger radii and lower temperatures (see Section \ref{sec: intrinsic UV excess}).

We note that the relationship between the stripped star and its companion mass, and thus the optical and UV flux ratio, are highly dependent on the assumptions made in our binary synthesis models, including the mass transfer and CEE prescriptions. We will further explore this in Section \ref{sec: dependence on underlying population}.

\subsection{Sources lost due to photometry}\label{sec: sources lost due to photometry}

In our forward model, we fail to recover UV photometry for $(54\,\pm\,1)\,\%$ (SMC) and $(74\,\pm\,2)\,\%$ (LMC) of our stripped stars. These sources are: (i) located outside of the SUMS footprint, or (ii) lack good UV photometry due to crowding.

\subsubsection{Survey coverage}\label{sec: footprint discussion}
We simulate stripped stars for an area of $\sim 25 \,{\rm deg}^2$ in the SMC and $\sim 95\,{\rm deg}^2$ in the LMC using the star formation histories derived by \cite{rubele_2018} and \cite{Mazzi_2021}. Only the central regions of each Cloud are surveyed by SUMS. We find that $\sim 30\,\%$ (SMC) and $\sim 50\,\%$ (LMC) of the simulated stripped stars lie outside the survey footprint and thus cannot be recovered by SUMS. In other words, our simulations predict that up to half of the stripped stars lie in the wings of the Clouds. This is broadly consistent with expectations from \citetalias{ludwig_2025}, who estimated that the SUMaC footprint covers $30-40\,\%$ of the recent ($<100$ Myr) star formation in the MCs.

We also typically lose $\lesssim 5$ sources across both Clouds due to masking of bright sources (Section \ref{sec: source injection}).

\subsubsection{Crowding}\label{sec: crowding}
Crowding is one of the main observational factors that impacts the completeness of the SUMS photometric catalog. As noted in \citetalias{ludwig_2025}, even though \textit{the Tractor} has improved performance over standard aperture photometry routines used by \textit{Swift}, because the UVOT PSF is slightly undersampled in the SUMaC images, \textit{the Tractor} cannot always properly disentangle the light of nearby sources. \citetalias{ludwig_2025} performed an initial quantification of this effect by injecting two sources into simulated UVOT images, moving them progressively closer together, and testing how well the injected magnitudes were recovered. This test was performed for three different pairs of brightnesses for the two sources. Overall \citetalias{ludwig_2025} found that \textit{the Tractor} performed well for neighboring sources located $>2''-2.5''$ away from each other, but that for sources with separation of $<1''-1.5''$, systematic errors could be introduced in the photometry. This was particularly the case for the fainter source in the pair, and \citetalias{ludwig_2025} therefore noted that caution should be taken when interpreting photometry from \textit{the Tractor} for sources with close neighbors-- especially if they are not the dominant source of light in the region. The benefit of our simulation is that we can now directly test (i) what fraction of stripped-star binaries crowding impacts and (ii) the response of the photometry to more complex/realistic stellar environments. These are critical to accurately assess the selection function.

\begin{figure*}
    \centering
    \includegraphics[width=\linewidth]{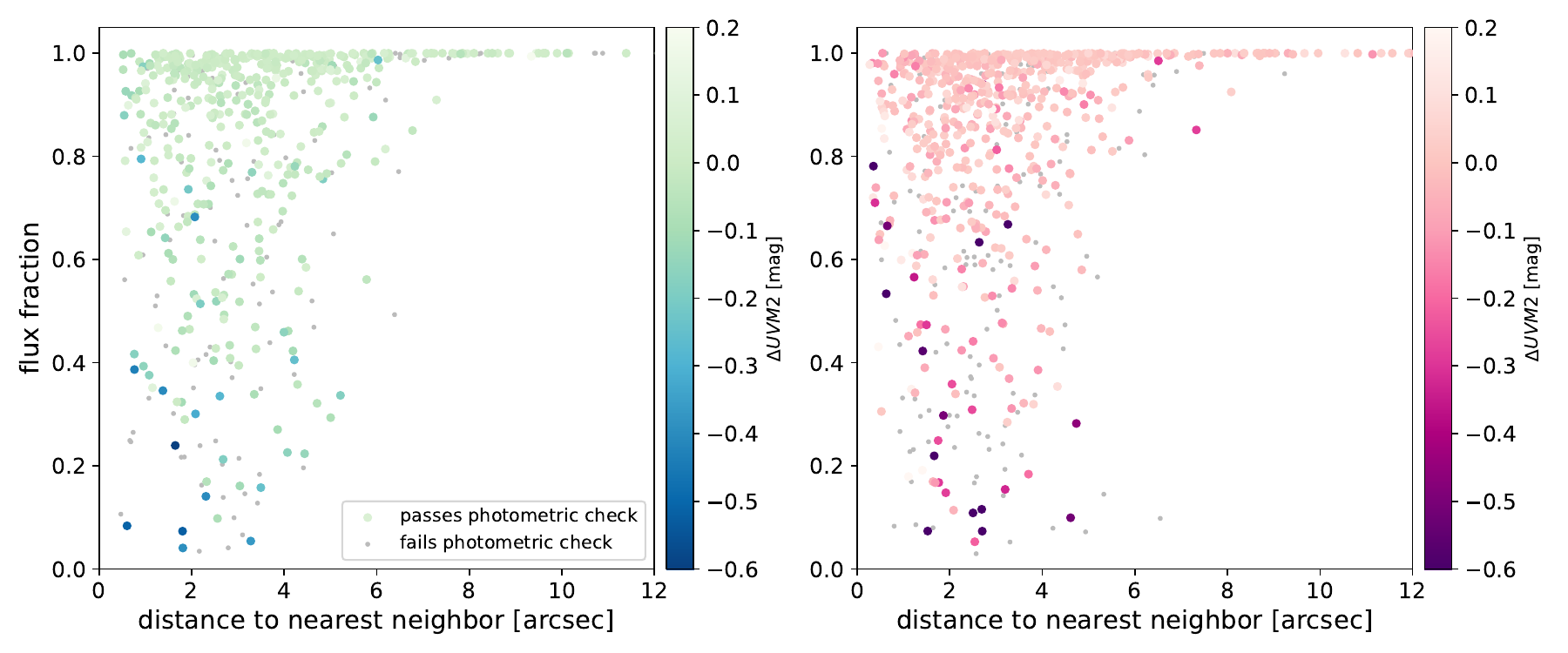}
    \caption{Average flux fraction as a function of distance to the nearest source. Gray markers represent injected sources which we recover mock photometry but fail the photometric quality checks. Colored markers show sources with robust photometry, with colors indicating the difference between the injected and recovered $UVM2$ magnitudes (i.e., $\Delta UVM2 = UVM2_{\rm{tractor}} - UVM2_{\rm{injected}}$ where negative values indicate that the UV flux has been overestimated). Simulated sources with overestimated UV fluxes from \textit{the Tractor} typically have low flux fractions ($< 0.5$) and lie within $3''$ of their nearest neighbor.}
    \label{fig:crowding_check}
\end{figure*}

We find that roughly $15\,\%$ of our injected sources lie within $1''$ (approximately the MCPS PSF) and roughly $65\,\%$ lie within $2.5''$ (approximately the SUMaC PSF) of an unrelated MCPS source. This implies that some stripped stars either will not be detected at all, or will have biased photometry, due to crowding. Figure \ref{fig:crowding_check} shows the difference between the injected and recovered magnitudes for the simulated stripped stars as a function of flux fraction and distance to the nearest neighbor.

We find that sources with their nearest neighbor located more than $>5''$ away typically have much higher flux fractions ($\gtrsim 0.9$). For injected sources with closer neighbors, we find a much larger range of flux fractions, extending down to 0. For sources in either Cloud, of those with robust UV photometry (i.e., those passing the photometric quality checks in Section \ref{sec: photometric quality cuts}), over $85\,\%$ have photometry recovered by \textit{the Tractor} that is within $0.2$ mag of the injected magnitude. These sources tend to cluster at higher flux fractions and are typically located farther from their nearest neighbors. The remaining $15\,\%$ of sources show larger discrepancies (e.g., $0.5-2$ mag) between recovered and injected magnitudes; they typically have low flux fractions $(< 0.5)$ and a neighboring source within $3''$.
 
Using the photometric pipeline from \citetalias{ludwig_2025}, we fail to recover any UV photometry for $(18\,\pm\,1)\,\%$ and $(33\,\pm\,3)\,\%$ of simulated stripped stars in the SMC and LMC that lie within the SUMS footprint (Section \ref{sec: mock photometry on images} and \ref{sec: combining mulitple measurements}). These sources are flagged as suspect in the post-processing of \textit{the Tractor} outputs. We reject photometry from \textit{the Tractor} for sources where the HEASARC calibration returns a flag (i.e., `SSS,' `EDGE,' or `SATURATED'), the magnitude uncertainty is greater than 0.36 mag, which indicates $<3\sigma$ detection, the residual fraction is greater than 0.3, or \textit{the Tractor} coordinates deviate by more than 1'' from the input coordinates. 
Before assessing UV excess, we require that simulated sources have at least two UV photometric measurements
(see Section \ref{sec: photometric quality cuts}). This further rejects $(15\,\pm\,1)\,\%$ and $(18\,\pm\,3)\,\%$ of the sources with photometry measured by \textit{the Tractor} in the SMC and LMC, respectively. All of the photometric quality cuts from \citetalias{ludwig_2025} are applied to flag sources that may 
lack robust UV detections. Given the high stellar density in the MCs, we suspect that most of these quality issues are related to crowding. \emph{The above statistics suggest that $\sim30\,\%$ (SMC) and $\sim45\,\%$ (LMC) of the systems with intrinsic UV excess within the SUMS footprint are missing from the final stripped star candidate catalog because crowding and survey depth prevent the measurement of UV photometry.}

Across both Clouds, up to $15\,\%$ of the sources that pass all quality cuts have observed UV photometry that is heavily biased due to crowding. In the most extreme cases, the injected and recovered magnitudes differ by up to 2 mag. This can lead to synthetic sources with mock observed fluxes that are (i) brighter than the true flux, leading to overestimated UV excess, or (ii) dimmer than the true flux, leading to underestimated UV excess. Our simulations suggest that about $5\,\%$ of the originally non UV excess stripped stars make it into our final simulated
candidate catalog due to overestimation of UV fluxes via
\textit{the Tractor}.

This raises a concern for contamination of the
stripped-star catalog by MS stars, which do not intrinsically
have UV excess but can appear to have UV excess if
the UV flux is overestimated by \textit{the Tractor}. We will explore
this further in Section \ref{sec: putiry}, when we forward model
a sample of MS stars to predict contamination rates.

\subsection{Extinction}\label{sec: extinction discussion}
\begin{figure*}
    \centering
    \includegraphics[width=\linewidth]{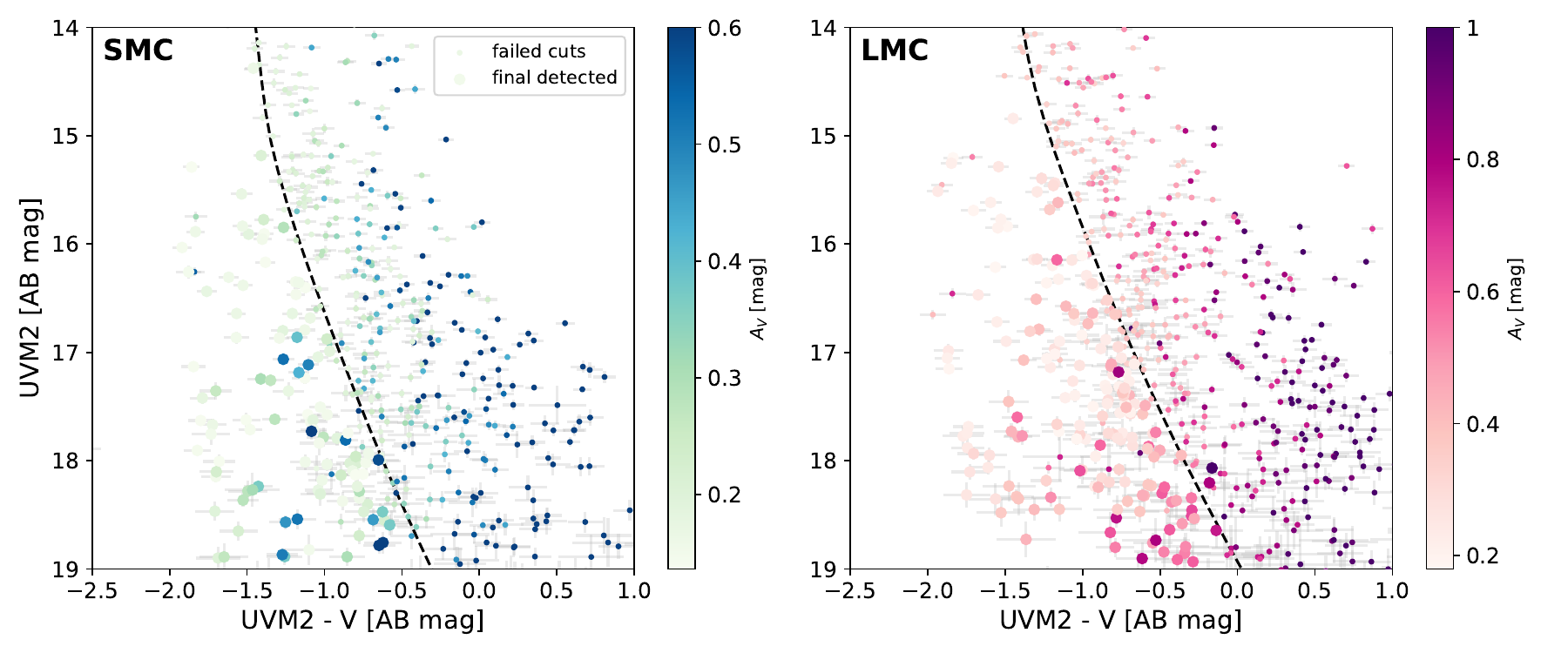}
    \caption{UV-optical CMD of the mock population for the SMC (left) and LMC (right). The photometry is de-reddened using a uniform extinction of $A_V = 0.22$ mag for the SMC and $A_V = 0.38$ mag for the LMC, following \citetalias{drout_2023} and \citetalias{ludwig_2025}. The marker colors indicate extinction. The larger markers represent sources in the final UV-excess sample. }
    \label{fig: mock cmd extiction}
\end{figure*}

\begin{figure}
    \centering
    \includegraphics[width=\linewidth]{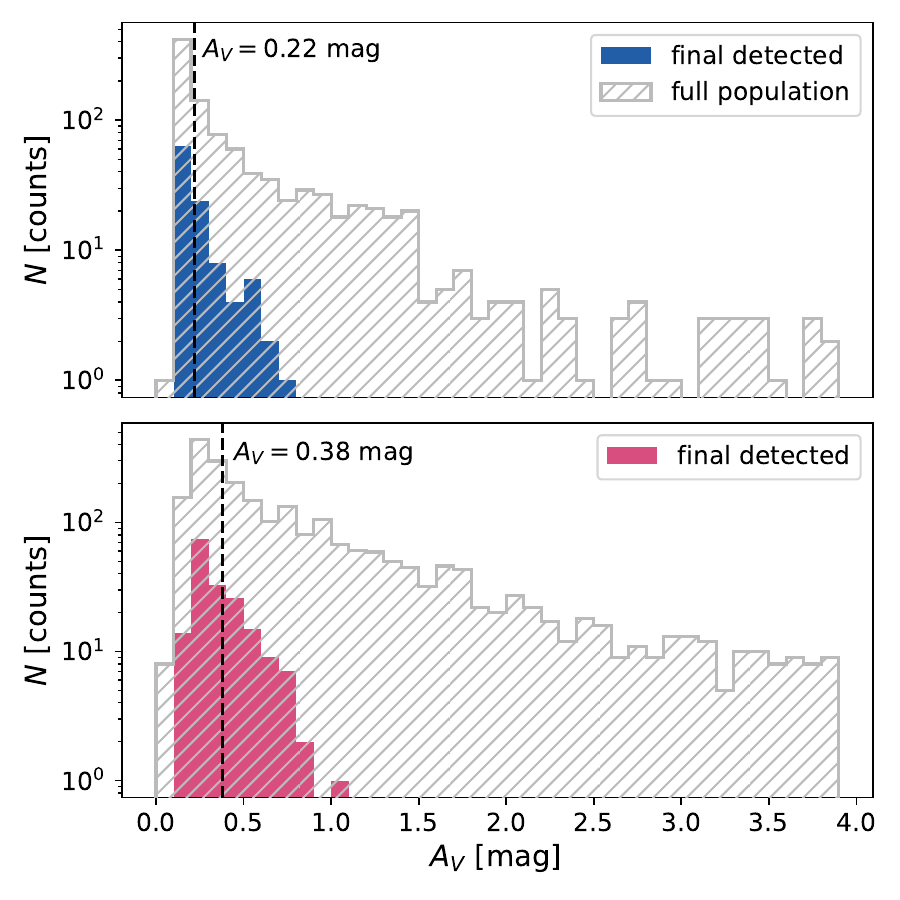}
    \caption{Distribution of reddening values for the synthetic stripped star population (top: SMC, bottom: LMC). The gray shaded histogram shows $A_V$ for the entire synthetic population, and the blue/pink histogram shows the final UV-excess sample. The median reddening of the full simulated population is $A_V \sim 0.26$ mag for the SMC and $A_V \sim 0.59$ mag for the LMC. The dashed line marks the the uniform extinction correction applied by \citetalias{drout_2023} and \citetalias{ludwig_2025}.}
    \label{fig: extinction histogram}
\end{figure}

Figure \ref{fig: mock cmd extiction} shows one of the UV-optical CMDs with sources colored by extinction.
Unsurprisingly, bluest sources have the least extinction while the reddest sources have higher extinction. The majority of sources that lie bluewards of the ZAMS have smaller extinction compared to those that lie redward of the ZAMS. Sources in the final UV-excess sample typically have $A_V \lesssim 0.5$ mag. In Figure \ref{fig: extinction histogram}, we plot the distribution of extinction values.
$\sim 55\,\%$ of the synthetic stripped stars in the SMC and $\sim 67\,\%$ in the LMC have reddening larger than the average reddening correction used by the \citetalias{ludwig_2025} UV-excess survey. This suggests the uniform correction assumed in SUMS underestimates the effects of reddening for the majority of the observed population.

In both Clouds, we see a sharp drop-off in the detection rate for sources with $A_V\geq  0.5$ mag and no UV-excess detections for sources with $A_V \geq 1$ mag. In the SMC, $45\,\%$ of the stripped stars with intrinsic UV excess (corresponding to $\sim 14\,\%$ of the total population) have $A_V \geq 0.3$ mag, and are missing from the final UV-excess sample due to underestimated extinction. Similarly, our simulations predict that roughly $60\,\%$ of the intrinsic UV excess population (corresponding to $\sim 26\,\%$ of the total population) are missing from the final detected sample due to having $A_V \geq 0.45$ mag. There is no strong correlation between stripped star mass and reddening.

We find that a small number of simulated candidates with high extinction ($A_V = 0.5 - 1$ mag) are recovered via UV excess. The majority of these are stripped-star binaries with compact object companions. Due to their large intrinsic UV excess, the bluest of these sources can exhibit UV excess even in relatively dusty regions. A small minority of candidates with high extinction have MS companions. These sources typically have bad \textit{Tractor} photometry due to being injected near a real source in the \textit{Swift}-UVOT image. This can lead to overestimated UV flux by \textit{the Tractor}, which can make the sources appear bluer than they are intrinsically. Most are flagged during the quality cuts, but a small fraction survive (see Section \ref{sec: crowding}).

\subsection{Mass distribution}\label{sec: mass /period distribution}
\begin{figure*}
    \centering
    \includegraphics[width=\linewidth]{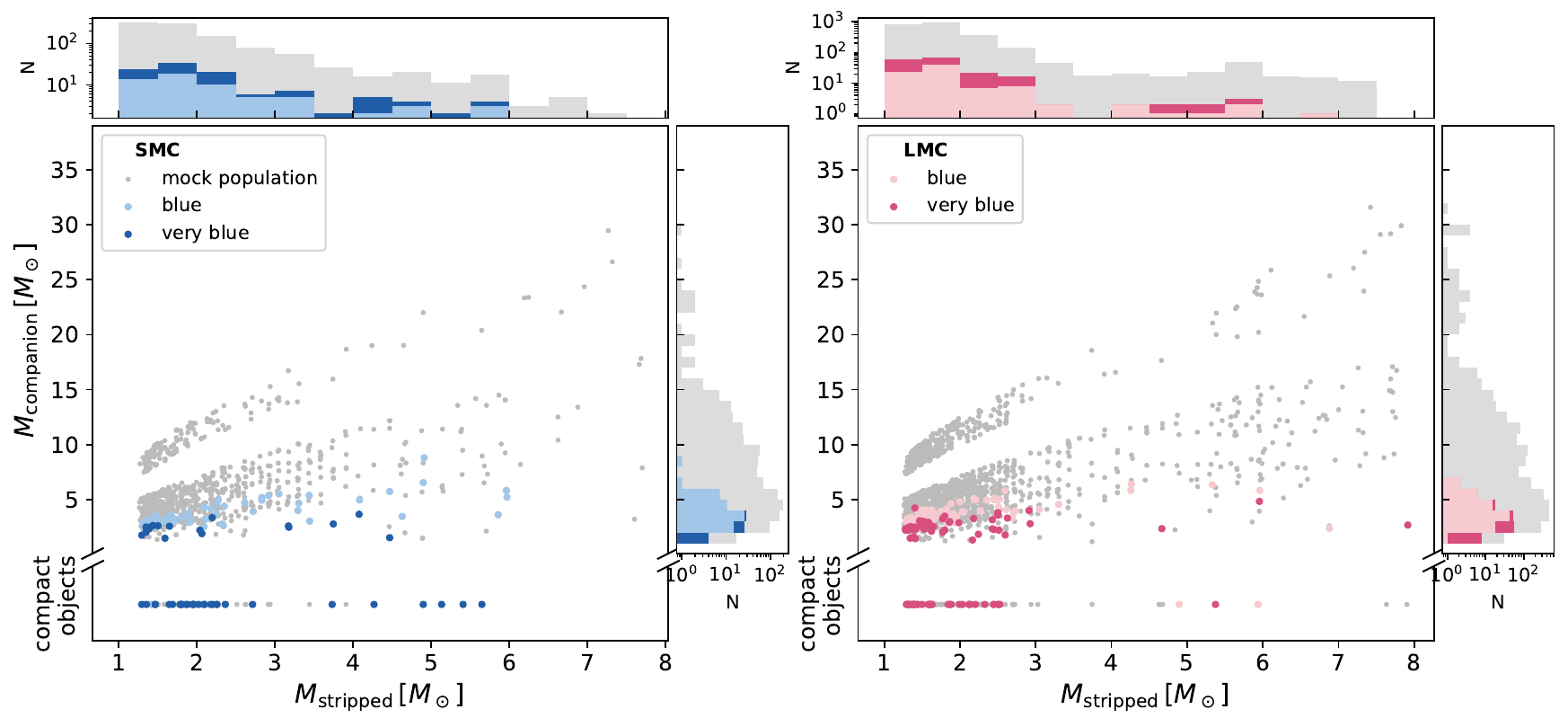}
    \caption{Stripped star and companion mass distribution for the synthetic population in the SMC (left) and LMC (right). For the compact object companion systems, we only show the stripped star mass, since we do not model the properties of the companion in our binary synthesis (Section \ref{sec: stripped stars with dark companions}). Grey markers show the entire synthetic population. Colored markers show sources that pass all candidate selection criteria. The UV-excess search is more sensitive to lower-mass stripped stars and companions. The typical stripped stars that pass all selection criteria are $M_{\rm stripped} \lesssim 5\,M_\odot$ and host low-mass MS companions ($M_{\rm MS}\lesssim 5\,M_\odot$).}
    \label{fig: mass period distribution}
\end{figure*}

Figure \ref{fig: mass period distribution} shows the stripped star and companion mass distribution of the simulated population. We only show the companion masses for MS companions since we do not compute masses for compact object companions (Section \ref{sec: stripped stars with dark companions}).

We find that all detected stripped stars have relatively low-mass MS companions, with no stripped stars hosting $M_{\rm MS} \gtrsim 10\,M_{\odot}$ companions detected in either Cloud. Even among the most massive intermediate-mass stripped stars ($M_{\rm{stripped}} \sim 7-8\,M_\odot$), only a few with MS companions are recovered in our forward model. Although massive stripped stars are intrinsically brighter, our results suggest that the detection efficiency of the search peaks at $M_{\rm stripped} \sim 2\,M_\odot$. This likely reflects the correlation between stripped star and companion masses in our model. The results of our simulations are consistent with the predictions of \citetalias{drout_2023}, who predict (their Figure S8) that UV excess is typically detected for stripped-star binaries with $M_{\rm MS} < (1.4-2.5) M_{\rm stripped}$.

The stripped-star mass is predicted to be correlated with its age. Thus, an upper bound on the detectable stripped-star mass allows us to constrain the age of the observed stripped stars. Stripped stars less massive than $\sim 5\,M_\odot$ are predicted to be at least $\sim 10$ Myr old since envelope stripping, while stripped stars less massive than $\sim 2\,M_\odot$ are predicted to be least $\sim 50$ Myr old \citep{hovis_2025}. This suggests that the majority of the observed stripped stars from SUMS are likley at least $(10-50)$ Myr old. We expect fewer stripped stars younger than $(5 - 10)$ Myr, because they are more likely to host massive companions, which dilute the UV excess of the stripped star.

\subsection{Comparison to the SUMS candidates}\label{sec: comparison to observed systems}
Below, we compare our simulated population to the photometric stripped star candidates from SUMS \citepalias{ludwig_2025}.

\subsubsection{Spatial distribution}
\begin{figure*}
    \centering
    \includegraphics[width=\linewidth]{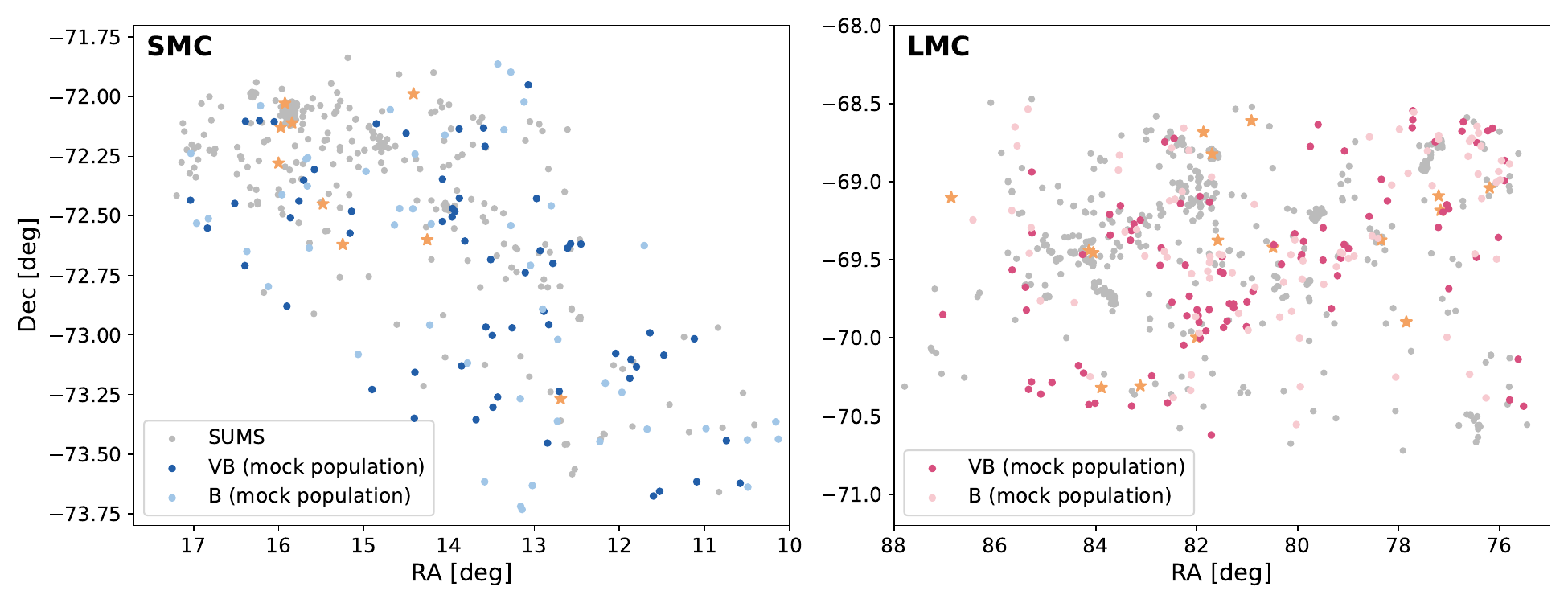}
    \caption{Spatial distribution of the stripped-star candidates in the SMC (left) and LMC (right). The gray markers show all SUMS photometric candidates (both ``Blue'' and ``Very Blue''), and the orange stars show the subset that has been spectroscopically confirmed \citepalias{drout_2023}. The blue/pink markers show the mock observed stripped stars, with the darker colored markers representing the ``Very Blue'' systems. In both Clouds, we find that observed candidates are more clustered compared to the simulated sources, which are more uniformly distributed.}
    \label{fig: spatial plot}
\end{figure*}

Figure \ref{fig: spatial plot} shows the spatial distribution of our forward-modeled population and the SUMS photometric candidates.
We find that the synthetic population is more smoothly distributed than the observed candidates, which appear clustered.

The clusters in the observed population trace regions smaller than the grid sizes of the star formation history from \citet{rubele_2018} and \citet{Mazzi_2021}, which we use to forward model the simulated stripped star population (see Section \ref{sec: stellar formation rate and syntehtic population}). The clusters are typically $(5-50)$ pc in diameter, while the subregions from the VMC survey have a diameter of $\sim 100$ pc. This suggests that these observed features may be smeared out in our simulated sample due to the spatial resolution of the star formation history.

Conversely, it is possible that these clusters are artificial features which are the results of spurious photometry within dense star clusters. These clustered regions contain a significantly larger fraction of faint UV sources ($\gtrsim 19$ mag) than the rest of the observed candidate population. One explanation for the overdensity of faint candidates is poor background subtraction due to excess light from nebulae surrounding star clusters. Imperfect background modeling can result in an overestimation of UV fluxes and thus spurious UV excess. For example, NGC 371 in the SMC -- which traces one of the largest clusters of the SUMS candidates -- is surrounded by a large H II region that is easily visible in broadband optical images. The cluster is much younger \citep[$5-10$ Myr; ][]{karoff_2008} than the typical age ($\sim 50$ Myr) of stripped stars we recover via UV excess (see Section \ref{sec: mass /period distribution}). This is particularly true for the faint ($\sim 19$ mag) sources, which would be low-mass stripped stars, which have longer life times ($\gtrsim 50$ Myr). This suggests a large fraction of the photometric candidates in these young star clusters may be false positives. Higher-resolution UV photometry (e.g., from {\it HST}) may be able to distinguish between these scenarios.

\subsubsection{Predicted magnitude distribution}\label{sec: compare uv photometry}
\begin{figure*}
    \centering
    \includegraphics[width=\linewidth]{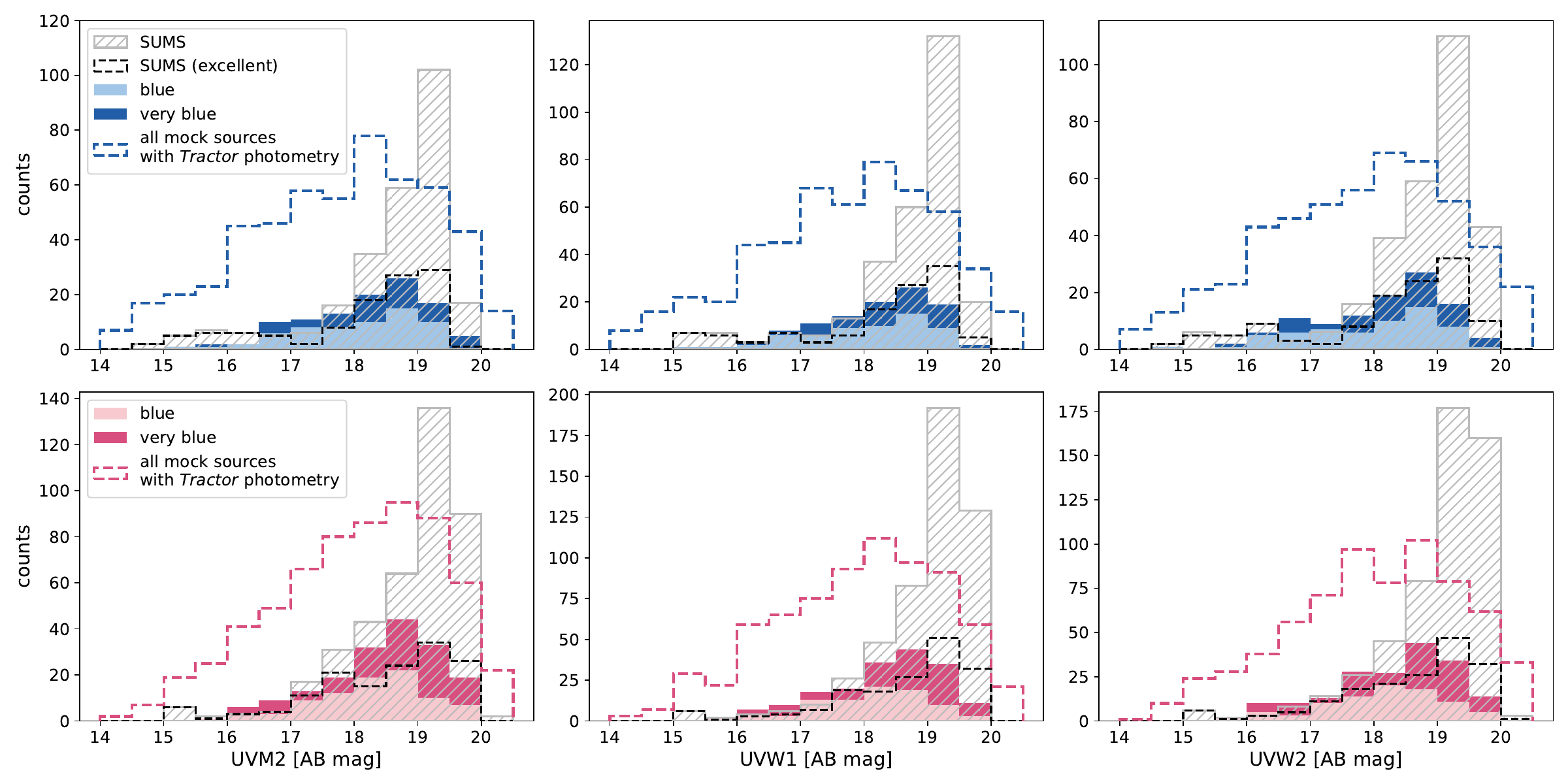}
    \caption{Magnitude distributions for each UV band measured by \textit{the Tractor}. The shaded gray histogram shows the magnitude distribution of the SUMS candidates. The dashed black line shows the SUMS candidates with ``Excellent'' rankings (i.e., `VB-E' and `B-E'). The solid blue/pink histogram shows the simulated candidates. The dashed blue/pink line shows all simulated stripped stars with mock observed photometry. The observed and simulated distributions agree in the intermediate magnitude range ($\sim 16-18$ mag). At fainter magnitudes, the SUMS catalog contains $5-10$ times the number of candidates than predicted by the simulated population.}
    \label{fig: UV magnitudes}
\end{figure*}

Figure \ref{fig: UV magnitudes} compares the distribution of UV magnitudes for our simulated sources and the observed stripped-star candidates.
In the intermediate magnitude range ($16$ mag $\leq UV \leq 18$ mag), the forward-modeled population reproduces the observed SUMS distribution well. This suggests that the model predictions of the $(2-5)\,M_\odot$ stripped stars are consistent with observations. Most spectroscopically confirmed stripped stars from \citetalias{drout_2023} also fall within this range. However, larger discrepancies between the observed candidates and simulated population are found at both brighter and fainter magnitudes.

For bright sources ($UV\lesssim16$ AB mag), we recover few to no candidates in the simulation, even though the underlying synthetic stripped star population (dashed lines in Figure \ref{fig: UV magnitudes}) contains them. In our model, these bright sources are typically massive stripped stars ($M_{\rm stripped} \gtrsim 6\,M_\odot$), which host relatively bright and massive companions, which dilute the UV excess of the stripped stars. This could imply that the brightest observed candidates are false positives. However, it is also possible that the masses of stripped stars and their companions are less strongly correlated in the observed sample compared to the synthetic population. This could occur, for example, if mass transfer is less conservative than assumed in our binary population synthesis model. We also caution that massive stripped stars, $M_{\rm stripped} \gtrsim 7\,M_\odot$, are not as well modeled in our population because we determine their properties through extrapolation (see Section \ref{sec: starburst parameters}).

At fainter magnitudes ($UV \gtrsim 19$ mag, corresponding to stripped stars with masses of $M_{\rm stripped}\sim 1.5\,M_\odot$ formed from $M_{\rm init} \sim 6\,M_\odot$), the SUMS catalog contains $5-10$ times more stripped-star candidates compared to our simulations, and roughly twice the stripped stars compared to the underlying synthetic population. This suggests several things: (i) the model significantly underpredicts the number of faint stripped stars, or (ii) a fraction of the faint observed candidates are other types of stars and thus are false-positive candidates. It is possible that adjustments to the underlying population model could alleviate some of this discrepancy. For example, we could increase the initial binary fraction rate for stars with $M_{\rm init} \sim 6\,M_\odot$, or use a non-uniform initial mass ratio distribution to increase the number of stripped stars with low-mass companions. However, we show below that the excess of faint sources is likely at least in part due to  contamination by young MS stars.

\subsection{Purity of the photometric catalog}\label{sec: putiry}
The purity of the SUMS catalog is uncertain, and as discussed in \citetalias{ludwig_2025}, one of the main expected contaminants is luminous MS stars (with other potential contaminants, including AGB stars and rapidly-accreting WDs, also discussed in \citetalias{ludwig_2025}). All MS stars by definition lie redward of the ZAMS; however, if a source has less extinction than the average $A_V$ assumed or has blended photometry that overestimates its UV flux, it can lie blueward of the ZAMS and exhibit UV excess. Below, we forward model a population of MS stars to test the level of contamination.

\begin{figure*}
    \centering
    \includegraphics[width=\linewidth]{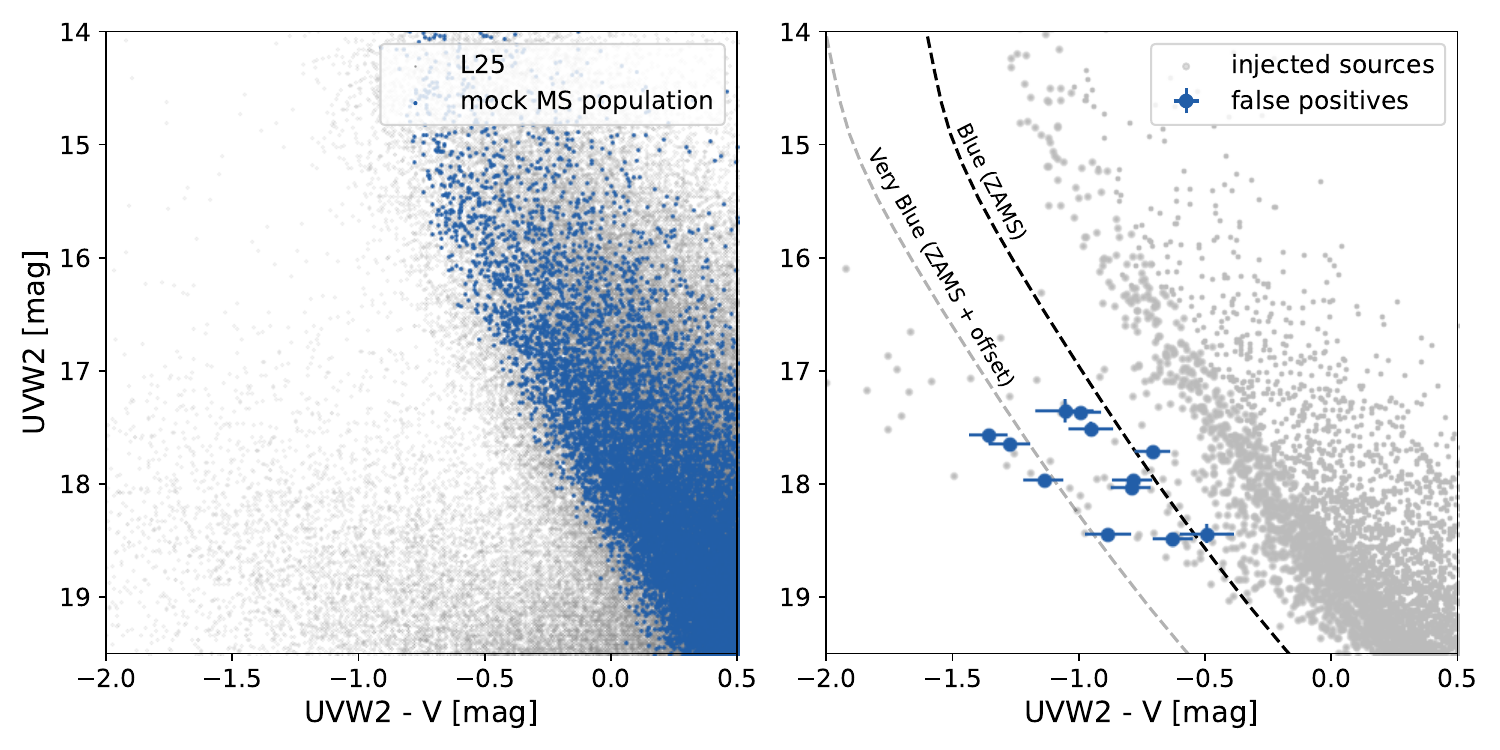}
    \caption{Left: UV-optical CMD with of \textit{Swift}-UVOT sources from \citetalias{ludwig_2025} (gray). In blue, we show the synthetic population of MS stars. The black dashed line shows the theoretical ZAMS and the gray dashed line shows the ZAMS offset by 0.4 indicating the ``Blue'' vs ``Very Blue'' cut off. Right: CMD of the 15,000 synthetic MS stars injected into \textit{Swift} images. Blue markers show the 12 sources that pass all quality cuts and thus would contaminate the UV-excess selected sample of stripped stars. The contaminant MS stars have significantly overestimated UV fluxes, mainly due to crowding. Such sources are likely to appear as stripped-star candidates in UV-excess searches. Our simulations predict $\sim 120$ MS contaminates, but this predicted contamination rate is sensitive to the adopted photometric corrections.}
    \label{fig: MS contaminants}
\end{figure*}

We construct a synthetic MS population for the SMC, using isochrones from MIST \citep{MIST_0, MIST_1} with a metalicity of [Fe/H] = -1.00. We use the ``Swift'' synthetic photometry to model the \textit{Swift}-UVOT photometry and the ``UBVRIplus'' synthetic photometry to model the MCPS photometry. We convert the UV photometry given in Vega magnitude to AB magnitudes using the conversions from Table \ref{table: AB vega conversion}. We assume an initial mass function from \cite{kroupa_2001} and the star formation history from \cite{rubele_2018}. We limit our population to MS stars formed in the past 500 Myr, since we expect stars detected in the \textit{Swift} images to be relatively young and massive. We assign a spatial coordinate and distance to each source using the same procedure described in Section \ref{sec: sptial distribition}, and assign extinction by integrating the dust models described in Section \ref{sec: extintion}. We inspect the UV-optical CMDs and find that the blue edge of the MIST synthetic MS is significantly bluer than that of the observed MS in the SUMS catalog. To improve the agreement between the shape of the observed and simulated MSs, we apply empirical corrections to the predicted photometry: $UVW2 = UVW2_{\rm MIST}  +0.3$, $UVM2 = UVM2_{\rm MIST} + 0.2$, $UVW1 = UVW1_{\rm MIST} +0.2$,  $V = V_{\rm MIST} - 0.2$, $I = I_{\rm MIST} -0.1$, and $B = B_{\rm MIST} - 0.25$. These corrections make the simulated MS stars redder than predicted by MIST, reducing the number of contaminants. In Figure \ref{fig: MS contaminants}, we plot a UV-optical CMD of the synthetic MS population and compare it to the UV photometric catalog from \citetalias{ludwig_2025}.

We randomly sample a subset of 15,000 stars from a population of $\sim$150,000 stars, which represent the total population of MS stars that lie inside the SUMS footprint and have $U < 20.5$ mag. We inject these sources into the \textit{Swift}-UVOT images and perform photometry using \textit{the Tractor} as described in Section \ref{sec: mock photometry on images}. In the right panel of Figure \ref{fig: MS contaminants}, we show one of the UV-optical CMDs of the injected sample of 15,000 MS stars. We apply the photometric and SED quality cuts described in Section \ref{sec: candidate selection} and require that sources exhibit UV excess in at least four of the nine UV-optical color combinations. Of the 15,000 MS stars, 12 pass all candidate selection criteria. Seven of the false-positive sources are classified as `B-G' and another two as `B-E.' Only three source have a `VB' color. With a contamination rate of $0.08\%$ for $\sim 150,000$ MS stars, our forward model predicts $\sim 120$ contaminants in the SMC.

\begin{figure}
    \centering
    \includegraphics[width=1\linewidth]{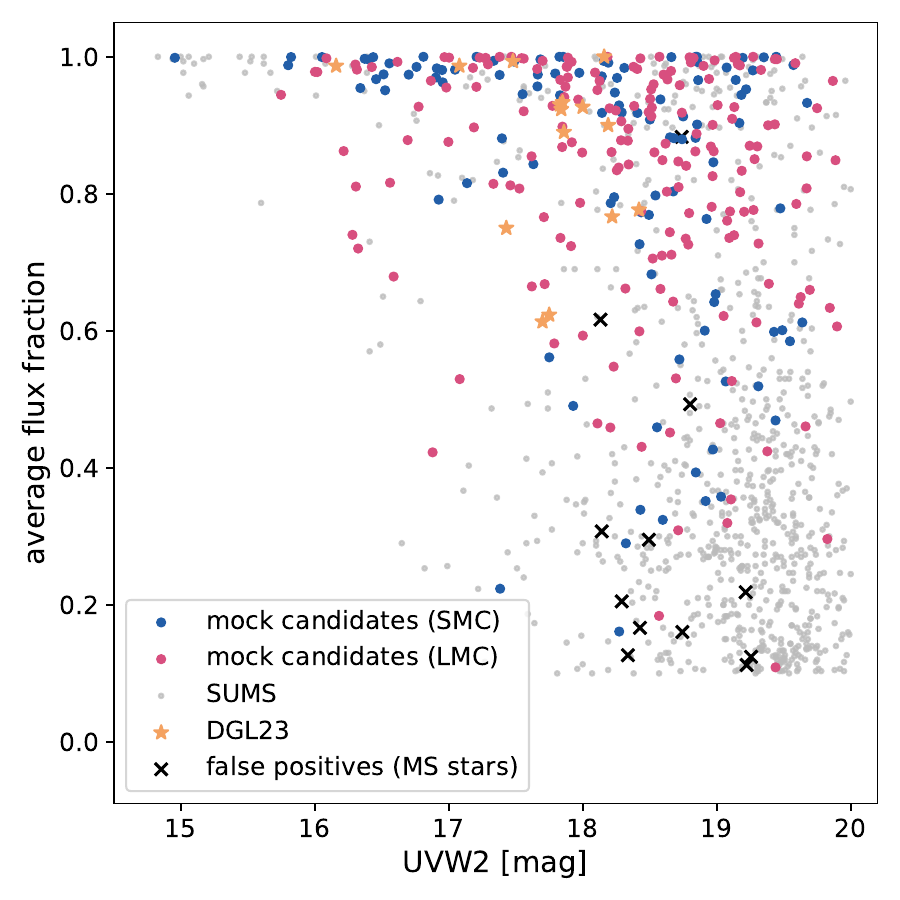}
    \caption{Flux fraction as a function of the $UVM2$ magnitude. Gray markers show all SUMS candidates. Orange stars show the class 1 and class 2 sources from \citetalias{drout_2023}. Colored markers show the simulated stripped stars in the SMC (blue) and LMC (pink) that make the final UV-excess sample. The observed candidates span a wide range of flux fractions, with some pileup near 0.1. The simulated stripped stars skew towards higher flux fractions. Black crosses show the 12 MS contaminants from the simulation outlined in Section \ref{sec: putiry}, which represent $\sim10\,\%$ of the total predicted false-positive population.}
    \label{fig: flux_fraction}
\end{figure}

We find that most of the false-positive sources have UV magnitudes reported by \textit{the Tractor} that are $>0.5$ mag brighter than the intrinsic brightness of the source. This is because the majority of these sources are injected into crowded regions and have a bright neighboring source within $3$ arcsec. Due to crowding, \textit{the Tractor} is unable to disentangle the flux of the two sources, leading to an overestimated UV flux. All of the simulated false-positive sources make the final UV-excess detection because of their overestimated UV fluxes, not because they fall in regions of low reddening. Figure \ref{fig: flux_fraction} shows the distribution of flux fractions for the observed candidates, simulated stripped stars, and simulated MS stars. A majority of the MS contaminants have relatively low flux fractions, reflecting the fact that they are located in crowded regions and have overestimated UV fluxes.

This possibility was recognized by \citetalias{ludwig_2025}. This motivated their flux fraction cut, in which all sources with flux fractions below 0.1 were removed, as well as those with flux fractions below 0.25 that have a neighboring source within $2.5''$. This criteria was designed to strike a balance between eliminating contaminants (i.e. the strong overdensity of SUMS candidates with low flux fraction in Figure \ref{fig: flux_fraction}; also see Figure 19 in Appendix of \citetalias{ludwig_2025}) while also keeping sources for future follow-up. While fewer than $10\,\%$ of the simulated stripped stars have flux fractions below 0.5, $\sim65\,\%$ of SUMS candidates have flux fractions below 0.5. Our simulations show that the excess of SUMS candidates with low flux fractions may be explained by false-positive candidates.

The number of predicted contaminants in the SMC ($\sim 120$) is roughly half the number of photometric stripped-star candidates from SUMS (298). This contamination rate is somewhat higher than the $\sim 15\,\%$ found in the \citetalias{drout_2023} spectroscopic sample. However, as discussed in Section \ref{sec: observed systems}, the preliminary version of the survey by \citetalias{drout_2023} used a different selection function, including manual selection of photometric candidates, which prioritized brighter sources with cleaner photometry than most of the photometric candidates. We find the inferred false-positive rate is sensitive to the photometric offsets that we apply to the MIST photometry, because the main selection of the UV-excess search relies on color measurements. When the photometric corrections are varied, the false-positive rate can vary significantly. For example, repeating the simulations without any photometric offsets, we estimate $\sim 2,400$ false positives in the SMC (which is significantly larger than the total number of photometric candidates that were found by SUMS). Thus, it is difficult to precisely predict the number of false positives in the SUMS catalog. However, the above experiment suggests that we can expect up to a few hundred false positives due to MS stars, and faint candidates ($> 18.5$ mag) with low flux fractions ($< 0.5$) are the most likely to be contaminants.

\section{Summary and Discussion}\label{sec: conclusion}
Recently, the first sample of intermediate-mass stripped stars in the MCs was discovered via a UV-excess survey (Figure \ref{fig: drout cmd}). This effort produced the SUMS catalog, which contains 820 photometric candidates \citepalias{ludwig_2025}. Of these, 16 have been confirmed spectroscopically \citepalias{drout_2023}. Our population synthesis model predicts that $\sim 1/3$ of stripped-star binaries show UV excess intrinsically. However, UV-excess searches for stripped stars are affected by several selection effects which decrease the survey completeness: (i) dust extinction and (ii) bright companions, which reduce the UV excess of stripped stars, and (iii) crowding and (iv) \textit{Swift} survey coverage, which limit the completeness of the UV photometry. In this work, we forward model a synthetic population of stripped stars to characterize the selection function of SUMS and constrain properties of the full population. Our main results are as follows:

\begin{enumerate}[label=(\roman*), wide, labelwidth=!, labelindent=0pt]
\item \textit{Stripped star population synthesis:} we create a synthetic population of intermediate-mass ($1 - 10\,M_\odot$) stripped stars in the MCs. By combining binary population synthesis with spatially-resolved star formation histories (Figures \ref{fig: LMC stripped star formation} and \ref{fig: mock stripped population spatial distribution}), we estimate the number, mass, and age distribution of stripped stars in each Cloud.  Our model predicts $\sim 1,000$ stripped stars in the SMC and $\sim 2,500$ in the LMC (Section \ref{sec: star formation rate}), which are comparable to predictions from other studies \citep[e.g., ][]{hovis_2025}. We spatially distribute simulated stripped stars so that their small-scale clustering follows that of young MS stars in the Clouds (Figure \ref{fig: mock stripped population spatial distribution}). We assume an exponential dust density distribution along the line of sight that is constrained by observations of the total integrated extinction column at each spatial location. This results in a realistic distribution of dust extinction, including both contributions internal to the Clouds and foreground extinction from the MW.

\item \textit{UV mock photometry:} we inject each synthetic stripped star into the \textit{Swift}-UVOT images used by SUMS and perform mock photometry using the photometric pipeline from \citetalias{ludwig_2025} (Figure \ref{fig: tractor exmaples}). By doing the injection and recovery, this allows us to model the selection biases from the finite spatial coverage of the \textit{Swift}-UVOT survey and blending of a stripped star's light with nearby sources in crowded regions. Using this pipeline, we recover $\sim 54\,\%$ (SMC) and $\sim 36\,\%$ (LMC) of our injected sources. We fail to recover photometry for the remaining sources because they are either: (a) located outside of the SUMaC footprint, (b) located in one of the masked regions near a bright source, or (c) removed in post-processing of \textit{the Tractor} outputs. We reject sources with bad photometry based on the selection criteria described in \citetalias{ludwig_2025}. Our finding suggests that $\sim10\,\%$ of stripped stars with robust intrinsic UV excess are missing from the final stripped-star catalog due to lack of good photometry alone.

\item \textit{UV-excess selection: }we apply the full SUMS selection criteria, including photometric and SED quality cuts, and UV-excess selection, to our simulated population (Figure \ref{fig: flowchart}). We find a stripped-star recovery rate of $(10.9\,\pm\,0.5)\,\%$ for the SMC and $(7.0\,\pm\,0.2)\,\%$ for the LMC (Figure \ref{fig:mock_pop_cmd}). Of the sources that show intrinsic UV excess, $(31.3\,\pm\,4.3)\,\%$ and $(15.3\,\pm\,0.6)\,\%$ of the simulated stripped-star systems are recovered; correcting for the SUMS footprint, this corresponds to a detection rate of $\sim47\,\%$ (SMC) and $\sim33\,\%$ (LMC) for the intrinsic UV-excess sources. The lower recovery rate in the LMC is likely a result of a combination of factors, including higher reddening due to dust and more crowding. Further, we classify our sources into four categories following \citetalias{ludwig_2025} (Table \ref{table: candidates}). Most of the synthetic stripped stars are classified as ``Excellent'' candidates, with similar numbers of sources classified as ``Blue'' and ``Very Blue.''

\item \textit{Completeness of the photometric catalog:} we predict that over $90\,\%$ of the intermediate-mass stripped stars in the MCs are not recovered by the UV-excess survey conducted by \citetalias{ludwig_2025}. The recovered stripped star population misses sources that: (a) host massive ($M_{\rm MS} \geq 5 - 7\,M_\odot$) MS companions (Figure \ref{fig: mock pop flux ratio}, \ref{fig: mass period distribution}), (b) lie in regions of high star density, or (c) have extinction of $A_V \geq (0.5-1)$ mag (Figure \ref{fig: mock cmd extiction}, \ref{fig: extinction histogram}). The simulations predict that such sources will not be selected by SUMS, even under ideal conditions. Stripped stars with compact object companions are more sensitive to the UV-excess survey since the companion does not dilute the stripped stars' UV excess. As a result, they are overrepresented in the detected population. While only $\sim10\,\%$ of intrinsic stripped stars host compact object companions in the model, we find that they make up about $22\,\%$ of the detected candidates. The overall recovery rate could be improved by carrying out surveys with higher spatial resolution or modeling extinction on a source-by-source basis (see Section \ref{sec: future surveys}).

\item \textit{Purity of the photometric catalog: }we compare the spatial distribution (Figure \ref{fig: spatial plot}) and UV magnitudes (Figure \ref{fig: UV magnitudes}) of the SUMS photometric candidates to our simulated population. 
In the observed sample, we find an excess of spatially clustered, faint photometric candidates ($UV \sim 19$ mag, which correspond to stripped stars of mass $M_{\rm stripped} \lesssim 1.5\,M_\odot$), that are not present in the synthetic population. We suspect that many of these sources are false positives due to (a) bad photometry in crowded regions, or (b) contamination by other types of UV-bright sources like young MS stars. We forward model a population of MS stars using the same methods we used for stripped stars to test for false positives (Figure \ref{fig: MS contaminants}). Most contaminants will likely have small UV flux fractions ($< 0.5$), faint magnitudes ($> 18.5$ mag), and moderate UV excess (Figure~\ref{fig: flux_fraction}). This implies higher contamination rates for faint candidates in crowded regions, which are typically classified as ``Blue'' candidates. The ``flux\_frac'' (the flux fraction) and ``dist\_neighbor'' (distance to the nearest neighbor) computed for the SUMS candidates in \citetalias{ludwig_2025} can be used to identify sources that have a higher likelihood of being contaminants and thus warrant greater caution.
\item \textit{Implications for stripped-star population synthesis and binary evolution}: We find that the binary synthesis models used in this paper \citep{Gotberg_2018} the observed intermediate-mass systems ($M_{\rm stripped} = 2-5\,M_\odot$) reasonably well but fail to match the higher-mass ($\gtrsim 5\,M_\odot$) and lower-mass ($\lesssim 2\,M_\odot$) ends of the observed population (Section \ref{sec: compare uv photometry}). This suggests that the assumptions adopted in our population models (Section \ref{sec: starburst parameters}) which we take as ground truth-- such as the treatment of mass transfer, overshooting, and CEE-- are broadly consistent with the observed population, at least in the intermediate-mass regime. The underprediction of high- and low-mass systems may be indicative of problems in the model but could also be a result of contamination of the observed candidates. Large-scale spectroscopic follow-up of the SUMS photometric candidates is necessary to distinguish between these possibilities.

\end{enumerate}

\subsection{Dependence of UV-excess detection rate on the underlying population}\label{sec: dependence on underlying population}
The recovery rate of stripped stars is sensitive to the assumptions made in modeling the underlying population. In our simulations, we take the binary population synthesis models described in Section~\ref{sec: starburst parameters} as ground truth. Below, we discuss how these assumptions affect the synthetic stripped star population and its detectability in UV-excess surveys.

Mass transfer efficiency is one of the most important factors in determining the detectability of stripped-star binaries via UV excess. However, it remains uncertain observationally, with a range of sometimes contradictory constraints presented in literature \citep[e.g., ][]{petrovic_2005, sun_2021, sun_2023, burt_2025, lechien_2025}.
Adjusting the mass transfer efficiency in our binary population synthesis models would alter the properties of the simulated stripped star population and could lead to different detection rates in a UV-excess survey. For example, a higher mass transfer efficiency is likely to result in brighter and more massive MS companions to stripped stars. Given our finding that a UV-excess search is most sensitive to lower-mass companions, a binary population produced with a higher mass transfer efficiency would likely produce a population which will have a lower detection rate. 

The overshoot parameter is another important factor in stellar evolution models that strongly influences the predicted properties of simulated stripped-star populations. For example, population synthesis models that adopt a smaller overshooting parameter than the one we use predict significantly few stripped stars \citep[e.g., ][]{Yungelson_2024}. We expect both the number and brightness of stripped stars to be  correlated with this parameter, and therefore predict that the detection rate will likely be sensitive to this value.

The initial binary population assumed in our stripped-star population synthesis is expected to affect the final predicted population. For example, a more bottom-heavy initial mass function or a higher binary fraction for low-mass primaries, could help alleviate some of the discrepancies between the observed population from SUMS and the synthetic population. We also assume that $7\,\%$ of stripped stars have compact object companions. Increasing this fraction would result in a higher detection rate of stripped-star binaries.

We also note that our underlying population does not account for single stripped stars, which are predicted theoretically \citep[e.g., ][]{Pols_1994_a, zhang_2012, Renzo_2019}. Observational evidence exists for single sdO/B stars \citep[e.g., ][]{Geier_2015, geier_2022} and such objects can be formed if a WD companion to a stripped star explodes as a Type Ia supernovae \citep[e.g., ][]{justham_2009}. We do not include single stripped stars in our synthetic population, since they are likely rare, but we note that they could be overrepresented in observed samples selected by UV excess.

Further work is needed to quantify the sensitivity of our predictions on detectability of stripped stars via a UV-excess survey. Empirical measurements of the SUMS detection rate will allow us to directly compare the predictions from various population models to the observed rates. Combining these results with constraints from sdB and WR star binaries, we can also test the mass dependence of binary interaction processes for systems formed through similar formation channels. Improved spectral model grids are necessary for further development of stripped star binary population synthesis models and work is currently in progress to address this. We plan to further investigate the sensitivity of our results to the underlying binary population synthesis in future work.

\subsection{Uncovering the missing population of stripped stars}\label{sec: future surveys}
Several improvements could be made to UV-excess searches to increase their completeness. 

Modeling extinction for individual sources via SED fitting, rather than assuming an average value for all sources, could increase the population detectable via UV excess. Our forward model predicts that about half the sources with robust intrinsic UV excess may be missed due to extinction. Precise measurements of reddening for individual sources might also allow for an extension of a similar survey to the MW, where reddening is more severe but 3D dust maps exist. 

Conducting a similar search using UV data from surveys with higher resolution or deeper photometry would allow us to significantly increase the number of stripped stars  detectable via UV excess. Current missions such as the Ultra Violet Imaging Telescope \citep[UVIT; ][]{UVIT}, and future missions like the Ultraviolet Explorer \citep[UVEX; ][]{kulkarni_2021}, are promising for follow-up studies and deeper explorations. These surveys also cover the FUV, which reaches shorter wavelengths than any of the \textit{Swift}-UVOT bands, making them well-suited for studying UV-bright sources. 

Lastly, to uncover stripped-star binaries with massive companions, a different search method, such as a spectroscopic search using surveys like SDSS-V \citep{Kollmeier2025} or light curve modeling \citep[e.g.][]{Naze2025} will be required.


\begin{acknowledgments}

We thank the referee for their constructive comments. We thank Hugues Sana's group at KU Lueven and Greg Green for useful discussions. This research was supported by NSF grants AST-2508988 and AST-2307232. BL has received financial support from the Flemish Government under the long-term structural Methusalem funding program by means of the project SOUL: Stellar evolution in full glory, grant METH/24/012 at KU Leuven. MRD acknowledges support from the NSERC through grant RGPIN-2019-06186, the Canada Research Chairs Program, the Ontario Early Researchers Award ER22-17-164, and the Dunlap Institute at the University of Toronto.

This project was developed in part at the Lorentz Center workshop ``From discovery to a population: benchmarking stripped stars and companions'' held in  2024 July in Leiden, Netherlands.

This research has made use of data and/or software provided by the High Energy Astrophysics Science Archive Research Center (HEASARC), which is a service of the Astrophysics Science Division at NASA/GSFC. This research has made use of the SVO Filter Profile Service  ``Carlos Rodrigo'', funded by MCIN/AEI/10.13039/501100011033/ through grant PID2023-146210NB-I00.

\end{acknowledgments}

%

\vspace{5mm}


\software{\textit{The Tractor} \citep{Lang_2016}, HEASARC \citep{heasarc}}, Astropy \citep{astropy:2013, astropy:2018, astropy:2022}



\appendix

\section{Estimating the dust disk scale height and inclination}\label{apdx: extinction calculation}
 
For each Cloud, we choose a small patch of the sky with significant extinction. Following Section \ref{sec: putiry}, we model a synthetic population of MS stars for the small patch. We then apply extinction to each of the synthetic MS stars based on the extinction model we describe in Section \ref{sec: extintion}. We vary the dust scale height $h_d$, and compare each of the resulting synthetic CMDs against the observed MCPS population from the same region of the sky. By comparing the observed spread and shape of the MS to the synthetic population, we estimate a dust scale height for each of the Clouds.

\begin{figure*}
    \centering
    \includegraphics[width=\linewidth]{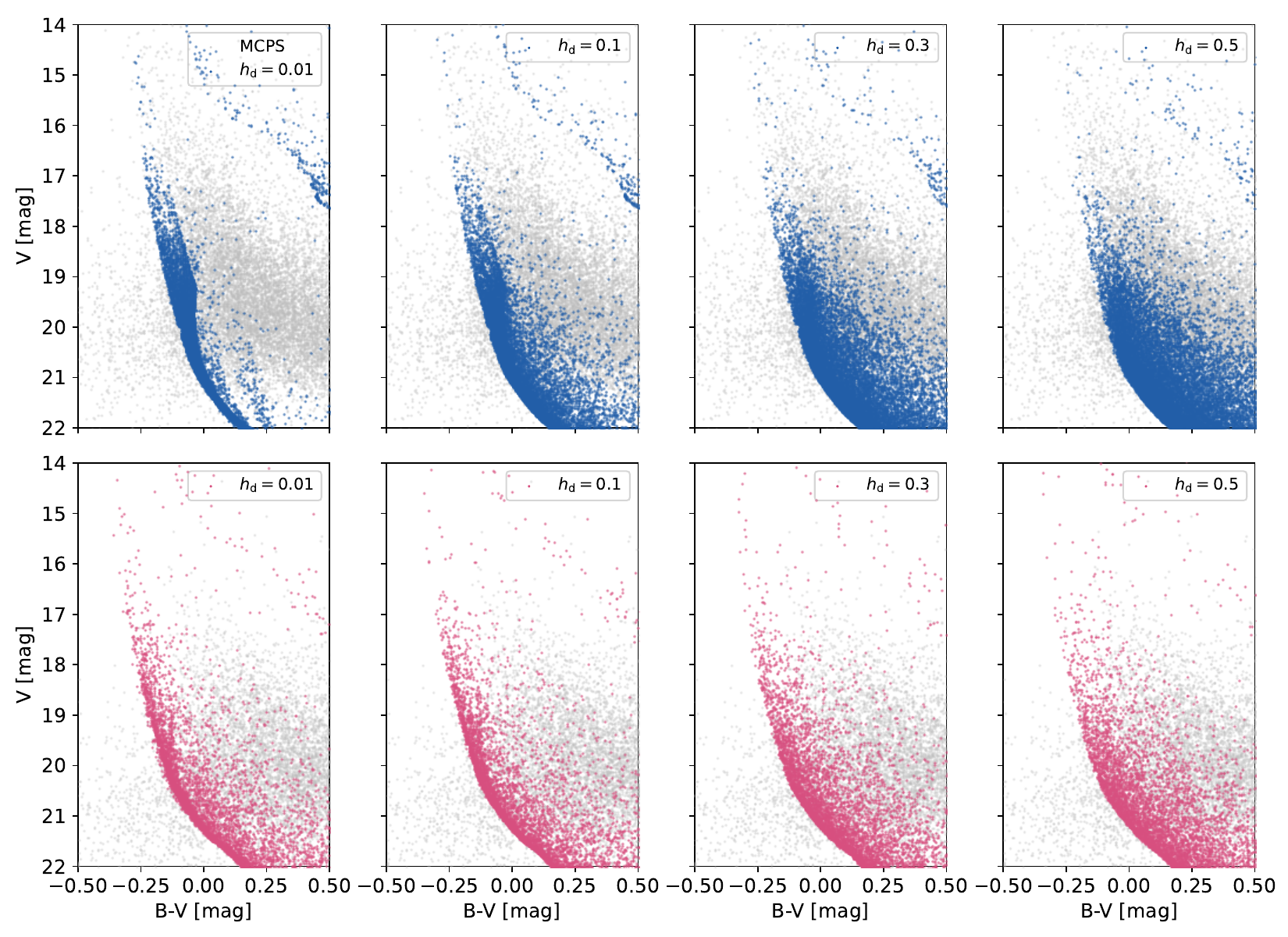}
    \caption{Colored markers show the synthetic MS population with extinction modeled using Equation \ref{eq: extinction}. Gray markers show the MCPS sources. The scale height is varied between each of the panels and is indicated in the legend. In both Clouds, the 
    synthetic MS and the MCPS MS show best agreement for a scale height of $h_{\rm d} \sim 0.3$.}
    \label{fig: dust scaleheight}
\end{figure*}

For the SMC, we restrict our analysis to $(11.8 \rm{\ deg} < \rm{RA} < 12.2 \rm{\ deg})$ and $(-73.3 \rm{\ deg}< \rm{Dec} < -73.05\rm{\ deg})$. The top panels of Figure \ref{fig: dust scaleheight} show some examples of the synthetic CMD with $h_{\rm d} = 0.01, 0.1, 0.3, 0.5$ kpc. We find that the synthetic and observed MS best agrees for $h_d \sim 0.3$ kcp. Since the geometry of the SMC is not well constrained we also repeat the analysis by applying an tilt to the disk (i.e., an inclination $i$), while fixing the scale height to $h_{\rm d} = 0.3$. We compare the synthetic and observed CMDs for $i =$ -30, -50, 0, 30, 50 degrees. We find that since we take the total integrated extinction at $z = \infty$ to be fixed, there is little to no correlation between inclination and the extinction map. Thus, we assume no inclination for the SMC.

For the LMC, we restrict our analysis to $(80.4$ deg $< \rm{RA} <80.6$ deg$)$ and $ (-69.8$ deg $ < \rm{dec} < -69.6$ deg$)$. We repeat the same analysis as above for the LMC to estimate the scale height. We find the best agreement between the observed and synthetic photometry for a scale height of $h_d = 0.3$ kpc. The bottom panels of Figure \ref{fig: dust scaleheight} show the comparison between the synthetic and MCPS CMD for the LMC.

\section{Results from several simulation trials}\label{apdx: multiple runs}
To assess the impact of random noise in our simulations, we repeat our complete forward modeling three times. In Table \ref{table: number of candidates}, we summarize the results for all simulations.
\begin{table}
\centering
\begin{tabular}{ c | c c c c c} 
 \hline
Galaxy & classification & run 1 & run 2 & run 3 & average\\ \hline\hline
SMC & VB-E  & 43 ($40\,\%$)&  42 ($39\,\%$) &  44 ($38\,\%$)&\\
  & VB-G  & 4 ($4\,\%$)&  8 (7$\,\%$) & 8 ($6\,\%$)&\\
& B-E  & 55 ($51\,\%$) &  52 (48$\,\%$) &  59 ($50\,\%$) &  \\
  & B-G  & 6  ($6\,\%$) &  6 ($6\,\%$) &  6 ($5\,\%$) & \\\cline{2-6}
 & total & 108 & 108 &  117 &111 \\
 & total simulated&1019 & 1052 &  997 &1023\\\cline{2-6}
 & detection rate &  $10.6\,\%$  & $10.3\,\%$& $11.7\,\%$ & $10.9\,\%$\\
 \hline
LMC & VB-E & 90 ($49\,\%$) &  76 ($44\,\%$) & 91 ($50\,\%$) & \\
  & VB-G  & 5 ($3\,\%$)& 8 ($5\,\%$)& 13 ($7\,\%$)& \\
& B-E & 85 ($47\,\%$)&  80 ($47\,\%$) & 73 ($40\,\%$)& \\
  & B-G & 2 ($1\,\%$) & 7 ($4\,\%$)& 4 ($2\,\%$)&\\\cline{2-6}
  & total &182 & 171  & 181 &  178 \\
  & total simulated & 2468& 2596 &2568 & 2544  \\\cline{2-6}
  & detection rate &  $7.3\,\%$ & $6.6\,\%$ & $7.0\,\%$ & $7.0\,\%$\\
 \hline
\end{tabular}

\caption{Results for the three separate forward-modeling trials outlined in Section \ref{sec: forward modeling}. The fraction of sources in each class is listed in parentheses. The classifications are as described in Section \ref{sec: candidate classification}.}
\label{table: number of candidates}
\end{table}
Our results are fairly consistent across all runs. The variance in the detection rate indicates an uncertainty of about $1\,\%$. Similarly, the fraction the mock candidates in each class (i.e., ``Very Blue'' vs ``Blue'' and ``Good'' vs ``Excellent'') is also consistent across runs, with an uncertainty of $3-5\,\%$.




\end{document}